\documentclass{article}

\usepackage{amsmath}
\usepackage[a4paper, margin=1in]{geometry}
\usepackage{graphicx}
\usepackage{natbib}
\usepackage{siunitx}
\DeclareSIUnit\year{yr}
\usepackage{hyperref}
\usepackage{bm}

\title{Towards reliable projections of global mean surface temperature}
\author{Philip G. Sansom, Donald Cummins, Stephan Siegert and David B. Stephenson}

\DeclareMathOperator{\dd}{d\!}                  
\DeclareMathOperator{\normal}{Normal}           
\DeclareMathOperator{\E}{E}                     
\DeclareMathOperator{\V}{Var}                   
\newcommand{\vect}[1]{{\bm{#1}}}                
\newcommand{\mat}[1]{{\bm{#1}}}                 

\newcommand{\data}{\vect{\mathcal{D}}}    
\newcommand{\out}{Y}                      
\newcommand{\obs}{Z}                      
\newcommand{\forcing}{X}                  
\newcommand{\latent}{\vect{\phi}}         
\newcommand{\params}{\vect{\psi}}         
\newcommand{\disc}{\delta}                
\newcommand{\shared}{\nu}                 

\newcommand{\state}{\vect{\theta}}        
\newcommand{\evolution}{\mat{G}}          
\newcommand{\forecast}{\mat{F}}           
\newcommand{\evolutionvar}{\mat{W}}       
\newcommand{\forecastvar}{\mat{V}}        
\newcommand{\evolutionreg}{\mat{D}}       
\newcommand{\priorexpstate}{\vect{a}}     
\newcommand{\priorvarstate}{\mat{R}}      
\newcommand{\postexpstate}{\vect{m}}      
\newcommand{\postvarstate}{\mat{C}}       
\newcommand{\priorexpobs}{\vect{f}}       
\newcommand{\priorvarobs}{\mat{Q}}        
\newcommand{\kalmangain}{\mat{K}}         

\newcommand{\cotwo}{CO$_2$}                  
\newcommand{\abrupt}{abrupt 4$\times$\cotwo}  

\begin{document}

\maketitle

\begin{abstract}
Quantifying the risk of global warming exceeding critical targets such as \SI{2.0}{\kelvin} requires reliable projections of uncertainty as well as best estimates of Global Mean Surface Temperature (GMST). 
However, uncertainty bands on GMST projections are often calculated heuristically and have several potential shortcomings. 
In particular, the uncertainty bands shown in IPCC plume projections of GMST are based on the distribution of GMST anomalies from climate model runs and so are strongly determined by model characteristics with little influence from observations of the real-world. 
Physically motivated time-series approaches are proposed based on fitting energy balance models (EBMs) to climate model outputs and observations in order to constrain future projections.
It is shown that EBMs fitted to one forcing scenario will not produce reliable projections when different forcing scenarios are applied.
The errors in the EBM projections can be interpreted as arising due to a discrepancy in the effective forcing felt by the model.
A simple time-series approach to correcting the projections is proposed based on learning the evolution of the forcing discrepancy so that it can be projected into the future.
These approaches give reliable projections of GMST when tested in a perfect model setting, and when applied to observations lead to well constrained projections with lower mean warming and narrower projection bands than previous estimates.
Despite the reduced uncertainty, the lower warming leads to a greatly reduced probability of exceeding the \SI{2.0}{\kelvin} warming target.
\end{abstract}

\section{Introduction}

Global Mean Surface Temperature (GMST) is a key quantity for projecting future climate since it integrates many large scale processes, and many changes and impacts scale with GMST \citep{IPCC2018,Suttonetal2015}.
It is also the summary measure most often used to communicate climate change to the public and to policy makers.
Credible assessments of the risk of global warming exceeding targets such as \SI{2}{\kelvin} set out in the Paris agreement are critical for policy makers to make informed decisions about how to meet those targets.
Therefore, it is important that the projections such assessments are based on are not only accurate, but also reliable, i.e., the probabilities of particular events are also accurate \citep{Broecker2012}.

Climate projections are usually derived from general circulation models (GCMs) designed to simulate the climate system as closely as is currently possible \citep{Collinsetal2013}.
Uncertainty in climate projections arises from many sources including imprecise initial conditions and natural variability \citep{Deser2012a}, the parameters of unresolved processes within a single GCM \citep{Collins2007}, choices made in constructing one GCM compared to another \citep{Tebaldi2007}, and uncertainty about future emissions.
Uncertainty about future emissions is usually addressed by conditioning projections on one or more predetermined scenarios \citep{Moss2010}.
Other uncertainties are usually quantified by analysing an ensemble of simulations that vary one or more of the uncertain factors.

The concept behind the use of ensembles for probabilistic projection is that the ensemble members represent samples from the distribution of plausible outcomes \citep{Smith2002,Palmer2006}.
In practice, limitations of the models, observations etc. affect the spread of the ensemble and reduce the skill of the forecast.
This is particularly problematic for multi-model ensembles which are not designed to span a space of possible model constructions, and are often referred to as ``ensembles of opportunity'' \citep{Knutti2010b,Stephenson2012}.
Multi-model projections of GMST in particular exhibit a very large spread of outcomes \citep{Collins2007}.
The IPCC approach to handling this uncertainty is to take anomalies with respect to a specified reference period \citep[Figures~T.S.14 \& T.S.15]{Collins2007}.
This reduces the spread of the projections in the future, but the projected warming and associated uncertainty then depend strongly on the reference period.
There is also no reason to believe the projections are probabilistically reliable.


Many methods have been proposed for reducing the uncertainty in multi-model ensemble projections by weighting models according to their past performance in simulating the observed climate \citep{Greene2006,Min2006,Bhat2011,Shiogama2011,Watterson2011}.
Some weighting methods have been shown to produce reliable projections of future climate in perfect model tests \citep{Abramowitz2015,Sandersonetal2017,Knutti2017,Strobach2020}.
However, others have questioned the use of weights based on past performance when projecting conditions that differ significantly from that past \citep{Stainforth2007,Sansometal2013b}.
The objection is that all models are fundamentally different from the system they represent and share common limitations that mean perfect model validation imparts only limited confidence.
Further, it has been shown that if the weights do not reflect the true model skill, then an unweighted ensemble may be preferred \citep{Weigel2010}.

The alternative to model weighting is to build a formal statistical framework relating climate models to the real-world \citep{Raisanen2001,Furrer2007,Buser2009,Annan2010,Annan2011}.
Some statistical frameworks effectively included performance weights of their own \citep{Tebaldi2005,Smith2009,Tebaldi2009}.
The most recent developments proposed independently by \citet{Chandler2013} and \citet{Rougieretal2013}, and extended by \citep{Sansometal2020} and \citet{Huang2019}, allow for common biases between climate models and the real-world due to common limitations of the models (e.g., missing processes, limited resolution, etc).
This can be thought of as separating model uncertainty (model differences) from model inadequacy (model limitations).

\begin{figure}[t]
  \centering\includegraphics[width=0.5\textwidth]{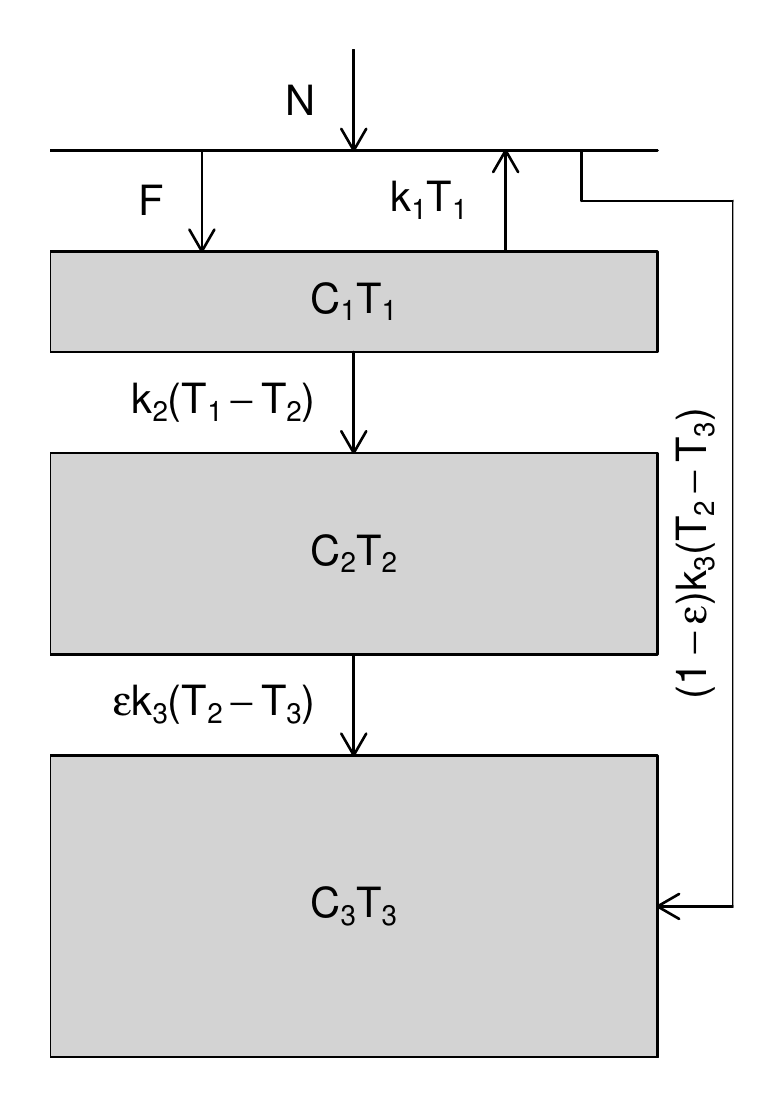}
  \caption{The 3-box EBM. The thickness of each box represents its heat capacity and the arrows indicate the flow of heat between boxes. The horizontal line represents the top of the atmosphere, which has no heat capacity.}
  \label{fig:figure2}
\end{figure}

\citet{Jonkoetal2018} proposed a statistical framework for projecting GMST where energy balance models (EBMs) are fitted to outputs from multiple climate models in order to form a prior distribution for the parameters of an EBM fitted to observations of the real-world.
Energy balance models are simple climate models that represent the atmosphere and ocean as a number of vertically stacked boxes, see Figure~\ref{fig:figure2}.
Fitting EBMs rather than purely statistical models has two main advantages: the model is physically motivated, and the parameters are physically interpretable.
\citet{Geoffroyetal2013a} found that the parameter estimates obtained from EBM fits to \abrupt\ experiments provided reasonable predictions of transient climate experiments where \cotwo\ is increased by $1\,\%$ per year, similar to the rate observed over the past 150 years.
However, \citet{Gregory2020} showed that estimates of key parameters are biased when fitted to historical observations rather than idealised climate experiments.
This implies that an EBM fitted to historical observations will lead to biased projections of future climate.
In this study, we show that fitting EBMs to idealised experiments will also lead to biased projections of future climate, and propose a physically motivated statistical approach to correcting the observed biases.
The proposed methodology produces probabilistically reliable projections when fitted to the historical period and tested in a perfect model setting.


The remainder of this study proceeds as follows.
In Section~\ref{sec:ebms}, we show that projections from EBMs fitted to idealised experiments are biased.
Section~\ref{sec:methodology} describes the proposed methodology, the statistical model for the forcing discrepancy, the relationship between the climate models and the real-world, and the strategy for sampling the parameters and making projections.
Section~\ref{sec:data} describes the data used to project future GMST, and methods for inference and model checking.
Section~\ref{sec:results} describes the results, including cross-validation to assess reliability, the distribution of the ECS, projections of future GMST up to 2100, and the probability of meeting the Paris agreement.
We finish in Section~\ref{sec:conclusions} with concluding remarks.

\section{Energy balance models and reliability}
\label{sec:ebms}

\citet{Jonkoetal2018} fitted two-box EBMs, however \citet{FredriksenRypdal2017} and \citet{Cumminsetal2020} found that three-box EBMs provide a better fit to both models and observations.
The three-box model fitted by \citet{Cumminsetal2020} is described by the following set of ordinary differential equations
\begin{align}
  C_1 \frac{\dd T_1}{\dd t} & =
    F - k_1 T_1 - k_2 \left( T_1 - T_2 \right) + w_T(t) & 
    w_T(t) & \sim \normal \left( 0, \sigma_T^2 \right)
    \label{eqn:t10} \\
  C_2 \frac{\dd T_2}{\dd t} & = 
    k_2 \left( T_1 - T_2 \right) - \varepsilon k_3 \left( T_2 - T_3 \right)
    \label{eqn:t20} \\
  C_3 \frac{\dd T_3}{\dd t} & = 
    k_3 \left( T_2 - T_3 \right)
    \label{eqn:t30}
\end{align}
where $T_1$, $T_2$ and $T_3$ are the temperatures in each box, $C_1$, $C_2$ and $C_3$ are the heat capacities of the boxes, $k_1$, $k_2$ and $k_3$ are heat transfer coefficients, and $F$ represents external forcing (i.e., \cotwo).
The top layer $T_1$ is usually assumed to represent the surface temperature and is the only observed quantity.
The stochastic term $w_T(t)$ represents natural variability in surface temperature, and $\varepsilon$ is the so-called efficacy factor introduced by \citet{Heldetal2010} to represent variation in $k_1$ during periods of transient warming.
An EBM is a linear time-invariant system, so completely characterised by its step response.
Therefore, EBMs are best fitted to idealised experiments containing a step change in forcing, such as the \abrupt\ experiment specified as part of the CMIP5 design \citep{Taylor2012}.
However, energy balance models are over-parameterised, making them difficult to fit even to step change experiments.
This difficulty can be overcome by including measurements of the net downward radiation flux at the top of the atmosphere $N(t)$, in addition to surface temperature, to constrain $k_1$ via the following relation
\begin{align}
  N(t) & = F(t) - k_1 T_1(t) + 
    (1 - \varepsilon) k_3 \left[ T_2(t) - T_3(t) \right].
  \label{eqn:radiation0}
\end{align}
Note that measurements of $N(t)$ are only available when fitting to GCMs, \emph{not} when fitting to the real-world.
To allow the natural variability in $N(t)$ to differ from $T_1(t)$, \citet{Cumminsetal2020} model the forcing $F(t)$ as red noise \citep{Hasselmann1976} so that
\begin{align}
  \frac{\dd F}{\dd t} 
    & = -\gamma \left[ F - F_C \forcing_C(t) \right] + w_F(t) & 
    w_F(t) & \sim \normal \left( 0, \sigma_F^2 \right)
    \label{eqn:forcing0}
\end{align}
where $F_C$ is the net radiative forcing due to a doubling of the atmospheric \cotwo\ concentration and
\begin{align}
  \forcing_C(t) & = \frac{1}{\log (2)} 
    \log \left[ \frac{\text{CO}_2(t)}{\text{CO}_2(0)} \right]
  \label{eqn:co2}
\end{align}
where $\text{CO}_2(t)$ is the \cotwo\ concentration at time $t$ \citep{Geoffroyetal2013a}.

Although the EBM representation is defined in continuous time, we only have uniformly spaced discrete model outputs and observations of the surface temperature $T_1(t)$ and radiation balance $N(t)$.
\citet{Cumminsetal2020} showed that the EBM can be discretised and written in state space form as
\begin{align}
  \vect{\out}(t) & = \forecast_d \state(t) + \vect{v}(t) & 
    \vect{v}(t) & \sim \normal \left( \vect{0}, \forecastvar_d \right) 
    \label{eqn:forecast_eqn} \\
  \state(t) & = 
    \evolution_d \state(t) + \evolutionreg_d \vect{\forcing}(t) + \vect{w}_d(t) & 
    \vect{w}_d(t) & \sim \normal \left( \vect{0}, \evolutionvar_d \right)
  \label{eqn:evolution_eqn}
\end{align}
where  $\vect{\out}(t) = \left[ T_1(t), N(t) \right]^\prime$ for $t=1,\ldots,T$ are the data, $\state(t) = \left[ F(t), T_1(t), T_2(t), T_3(t) \right]^\prime$ is the state, and $\vect{\forcing}(t)$ is the forcing (see Appendix~\ref{app:disc} for details).
The stochastic term $\vect{v}(t)$ represents observation and measurement error which is set to zero for climate model output.
The discretisation is exact provided the external forcing is piecewise constant, i.e., $\forcing(t)$ is constant between times $t$ and $t+1$. 
Efficient maximum-likelihood estimation of the EBM parameters can be achieved by using the Kalman filter to evaluate the model likelihood (see Appendix~\ref{app:disc}).
Once the EBM parameters are estimated by fitting to \abrupt\ experiments, it is straightforward to make predictions for other scenarios by applying appropriate \cotwo\ forcing through $\forcing_C(t)$.
The maximum-likelihood fitting allows us to quantify not only projection uncertainty due to natural variability in forcing and temperature, but also uncertainty about the fitted parameters (see Appendix~\ref{app:proj} for details).
Figure~\ref{fig:reliability} shows the results of using the EBM fits of \citet{Cumminsetal2020} to project future GMST and top-of-atmosphere radiation balance using the equivalent \cotwo\ forcing for the CMIP5 historical and RCP4.5 scenarios \citep{Meinshausenetal2011}.

Figures~\ref{fig:reliability}(a) and (b) compare the EBM projections for the HadGEM2-ES model with output from that model for the historical and RCP4.5 scenarios.
The radiation balance looks reasonable, but there are clear biases in the GMST and the model output regularly exceeds the credible interval.
Figures~\ref{fig:reliability}(c) and (d) show the standardised prediction errors (see Appendix~\ref{app:proj}) for all 16 climate models fitted by \citet{Cumminsetal2020}.
If the EBM projections were probabilistically reliable, then approximately 95\,\% of the standardised errors should lie between $-2$ and $+2$ with good scatter between those bounds at all times (assuming the projections approximately follow a normal distribution).
The radiation balance projections appear fairly reliable, with the exception of a series of strong negative spikes affecting all models during the historical period, and a slight positive bias visible in the multi-model mean error.
However, the GMST projections are clearly not reliable, with standardised errors not only regularly but continuously exceeding six standard deviations, similar strong negative spikes during the historical period, and a positive mean bias that grows throughout the historical period.

\begin{figure}[t]
  \centering\includegraphics[width=\textwidth]{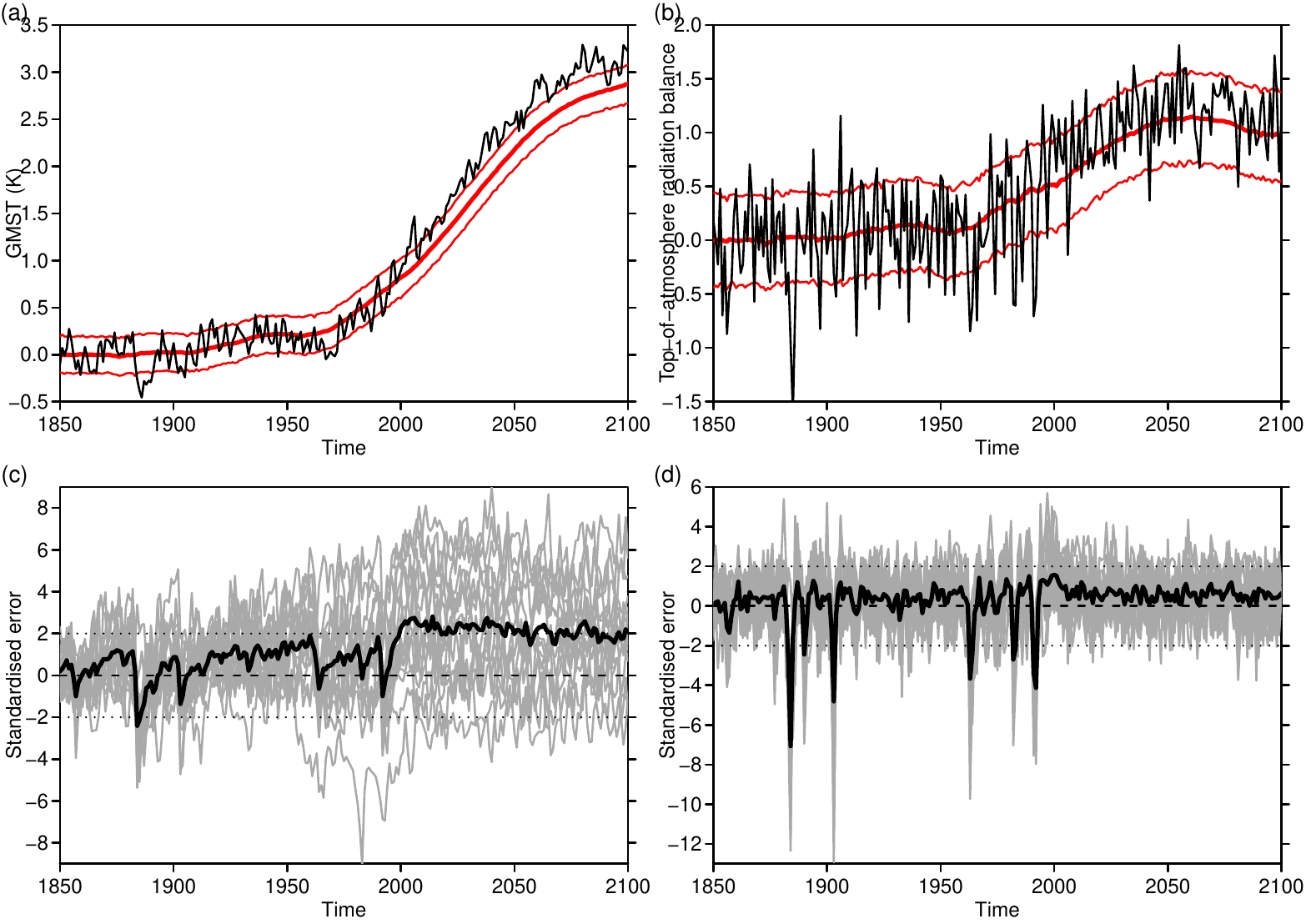}
  \caption{EBM projection reliability.
           (a) global mean surface temperature, and (b) top-of-atmosphere radiation balance in the HadGEM2-ES model under the RCP4.5 scenario.
           Black lines are the model simulations, thick red lines are the posterior predictive means based on fitting the EBM to the abrupt4xCO2 scenario, and thin red lines are a marginal 90\,\% credible interval.
           (c) and (d) Standardised prediction errors for global mean surface temperature, and top-of-atmosphere radiation balance respectively under the RCP4.5 scenario based on fitting to the abrupt4xCO2 scenario.
           Thin grey lines represent standardised errors from individual CMIP5 models.
           The thick black line is the ensemble mean.}
  \label{fig:reliability}
\end{figure}

The large standardised errors in Figures~\ref{fig:reliability}(c) and (d) indicate that the models are individually biased and over-confident, i.e. the credible intervals in Figures~\ref{fig:reliability}(a) and (b) are too narrow.
The large multi-model mean bias indicates that there are common biases affecting all models.
Neither the individual or common biases are surprising.
The EBMs are very simple linear representations of a much more complicated non-linear system.
There are many processes and feedbacks included in the climate models (and the real-world) that are not represented in the EBMs, e.g., albedo changes due to sea and land ice loss.
The effects of these processes will vary over time depending on the forcing scenario.
Consequently, fitting an EBM to any single scenario will result in compensating errors in the parameters.
Therefore, an EBM fitted to one scenario should not be expected to produce reliable projections when forced with a very different scenario.

One way to produce more reliable projections might be to fit to model outputs from the \abrupt, historical and future scenarios simultaneously in order to find a common set of parameters.
In practice, we found that this strategy results in parameters that do not fit any of the scenarios well.
Figure~\ref{fig:reliability} suggests an alternative strategy.
The spikes in both the radiation balance and GMST during the historical period in Figures~\ref{fig:reliability}(c) and (d) are due to volcanic eruptions injecting aerosols into the stratosphere, the effects of which are not included in RCP4.5 \cotwo\ equivalent forcings, i.e., a missing component of forcing.
Therefore, why not also treat the other biases as a discrepancy in effective forcing?
Figure~\ref{fig:reliability} and the EBM equations support this interpretation.
There are large errors and a large bias in GMST, but almost none in the radiation balance.
An effective discrepancy in forcing $F(t)$ directly affects both variables, but is balanced in the radiation balance equation by the negative $k_1 T_1(t)$ term acting to cancel out the discrepancy.
This is the approach we adopt in Section~\ref{sec:methodology}, we fit the EBMs to the \abrupt, historical and future scenarios simultaneously, allowing for stratospheric aerosol forcing and an effective forcing discrepancy in the historical and future scenarios.

%

\section{Towards more reliable projections}
\label{sec:methodology}

We propose a hierarchical Bayesian model to allow inference about future climate in the real-world by combining climate model outputs with observations of the real-world.
We assume that the output of each climate model, and the observations of the real-world can be represented by an EBM.
We combine surface temperature and top-of-atmosphere radiation outputs from climate models forced by an \abrupt\ scenario to learn the unique EBM representation of each climate model.
We combine surface temperature and top-of-atmosphere radiation outputs from the same climate models forced by historical and future climate experiments to learn the response of each model to volcanic forcing and the discrepancies visible in Figure~\ref{fig:reliability}.
We assume that the climate model outputs and hence the EBM parameters (basic and discrepancy) are exchangeable and arise from a common distribution which we also learn.
The common distribution over the model parameters provides a prior for the EBM representation of the real-world.
This prior distribution is critical in inferring the parameters for the real-world since only surface temperature observations are available, and only for the historical period.
Without the prior distribution obtained from the models, inference for the real-world would be almost impossible.

\subsection{Modelling the forcing discrepancy}

For the historical and future scenarios, we expand the EBM to include volcanic forcing $\forcing_V(t)$ and a forcing discrepancy $\disc(t)$ so that
\begin{align}
  \frac{\dd F}{\dd t} 
    & = -\gamma \left[ F - F_C \forcing_C(t) - F_V \forcing_V(t) \right] + w_F(t) & 
    w_F(t) & \sim \normal \left( 0, \sigma_F^2 \right)
    \label{eqn:forcing} \\
 C_1 \frac{\dd T_1}{\dd t} & =
    F + \delta - k_1 T_1 - k_2 \left( T_1 - T_2 \right) + w_T(t) & 
    w_T(t) & \sim \normal \left( 0, \sigma_T^2 \right)
    \label{eqn:t1} \\
  C_2 \frac{\dd T_2}{\dd t} & = 
    k_2 \left( T_1 - T_2 \right) - \varepsilon k_3 \left( T_2 - T_3 \right) 
    \label{eqn:t2} \\
  C_3 \frac{\dd T_3}{\dd t} & = 
    k_3 \left( T_2 - T_3 \right)
    \label{eqn:t3}
\end{align}
where $F_V$ is the radiative coefficient of volcanic forcing, and
\begin{align}
  N(t) & = F(t) + \delta(t) - k_1 T_1(t) + 
    (1 - \varepsilon) k_3 \left[ T_2(t) - T_3(t) \right].
  \label{eqn:radiation}
\end{align}

Figure~\ref{fig:reliability} indicates that we need to split the discrepancy $\disc(t)$ into shared and model-specific components.
If we try to model the shared and model-specific discrepancy components independently, then the statistical model is not identifiable since we have $M+1$ discrepancy components and only $M$ time series, where $M$ is the number of models.
Therefore, we model the forcing discrepancy in model $m$ as
\begin{align}
  \frac{\dd \disc}{\dd t} & = \shared(t) + w_\disc(t) & w_\disc(t) & \sim \normal \left( 0, \sigma_\disc^2 \right)
  \label{eqn:disc}
\end{align}
where 
\begin{align}
  \frac{\dd \mu}{\dd t} & = \shared(t) & \shared(t) & \sim \normal \left( 0, \sigma_\shared^2 \right).
  \label{eqn:shared}
\end{align}
So, in discrete time, each model-specific discrepancy $\disc(t)$ is modelled by a random walk about a common mean $\mu(t)$ which is itself modelled by a random walk. 
This parametrisation introduces two additional model-specific parameters $F_V$ and $\sigma_\disc$, and one shared parameter $\sigma_\shared$.
Note that we refer to $\mu(t)$ and $\nu(t)$ interchangeably as \emph{the shared discrepancy} since although $\mu(t)$ is the more interpretable quantity, it never actually enters the model formulation, so $\nu(t)$ is the object of inference.

\subsection{Learning the distribution over the models}
\label{sec:ensemble}

Each EBM representation has 13 model-specific parameters which we collect into a vector
\begin{align*}
  \vect{\latent}_m & = \left( \gamma, C_1, C_2, C_3, k_1, k_2, k_3, \varepsilon, \sigma_F, \sigma_T, F_C, F_V, \sigma_\disc \right)^\prime & \text{for} \ m & = 1,\ldots,M.
\end{align*}
We assume that the models are exchangeable, i.e., without prior knowledge about the performance of a particular CMIP5 model, we would specify the same prior beliefs about the parameters for every CMIP5 model.
All of the model-specific parameters in $\latent_m$ are constrained to be positive.
Therefore, it is convenient to model the relationship over the models as
\begin{align}
  log \left( \latent_m \right) & \sim \normal \left( \vect{\mu}_\latent, \mat{\Sigma}_\latent \right) & \text{for} \ m & = 1,\ldots,M
  \label{eqn:params}
\end{align}
where $\vect{\mu}_\latent$ is a real vector of length 13, and $\mat{\Sigma}_\latent$ is a $13 \times 13$ symmetric positive-definite matrix.

\subsection{Learning about the real-world}
\label{sec:realworld}

We assume that the real climate system can be approximated by an EBM identical to those used to represent the climate model outputs during the historical and future periods (Equations~\ref{eqn:forcing}--\ref{eqn:disc}).
The real climate system is assumed to have its own unique vector of parameters 
\begin{align*}
  \vect{\latent}_\obs & = \left( \gamma, C_1, C_2, C_3, k_1, k_2, k_3, \varepsilon, \sigma_F, \sigma_T, F_C, F_V, \sigma_\disc \right)^\prime.
\end{align*}
Following \citet{Rougieretal2013} and \cite{Sansometal2020} we assume that the real climate system is co-exchangeable with the climate models so that
\begin{equation}
  \log \left( \latent_\obs \right) \sim \normal \left( \vect{\mu}_\latent, \kappa^2 \mat{\Sigma}_\latent \right)
  \label{eqn:obs_par}
\end{equation}
where $\kappa$ is a positive real scalar.
This formulation implies that our expectation for the real-world is the same as the models, and the correlations between parameters are the same, but our uncertainty may be different due to missing processes and other errors in the models.
If $\kappa = 1$ then the real-world is assumed to be exchangeable with the models, i.e., just another climate model, no missing processes, etc.
Setting $\kappa > 1$ implies that given the knowledge gained from the models, we are less certain about how the real-world will behave than about how a new model would behave.
This seems appropriate given the course representation and the number of processes still missing from climate models.

\subsection{Projecting future climate}
\label{sec:projection}

Since the real-world is assumed to have the same EBM representation as the models during the historical and future periods, it also depends on the shared discrepancy $\shared(t)$ through Equation~\ref{eqn:disc}.
In Section~\ref{sec:inference} we outline how to sample both the real-world parameters $\latent_\obs$ and the shared forcing discrepancy $\shared(t)$ (see Supplementary Material for details).
Conditional on knowing $\latent_\obs$ and $\shared(t)$, the discretised form of Equations~\ref{eqn:forcing}--\ref{eqn:disc} can be written in state space form (Equations~\ref{eqn:forecast_eqn} and \ref{eqn:evolution_eqn}, see Appendix~\ref{app:full} for details), and the future climate of the real-world can be sampled as follows:
\begin{itemize}
  \item Sample $\latent_\obs^\star$;
  \item Sample $\shared^\star(t)$ for $t = 1,\ldots,\tau_H,\ldots,\tau_F$ where $\tau_H$ and $\tau_F$ are the ends of the historical/observed and future periods respectively;
  \item Use the Kalman filter (Equations~\ref{eqn:state_pred}--\ref{eqn:state_post}, Appendix~\ref{app:disc}) to compute $\left[ \state_\obs(\tau_H) \mid \data(\tau_H), \latent_\obs \right]$, i.e., the distribution of the state of the real-world at the end of the observed period. 
  \item Sample $\state_\obs^\star(\tau_H)$ from $\left[ \state_\obs(\tau_H) \mid \data(\tau_H), \latent_\obs^\star \right]$;
  \item For $t$ in $\tau_H+1,\ldots,\tau_F$
  \begin{itemize}
    \item Sample $\state_\obs^\star(t)$ from Equation~\ref{eqn:evolution_eqn}, conditional on $\state_\obs^\star(t-1)$, $\vect{\forcing}(t)$ and $\latent_\obs^\star$;
    \item Sample $\obs^\star(t)$ from Equation~\ref{eqn:forecast_eqn}, conditional on $\state_\obs^\star(t)$ and $\latent_\obs^\star$;
  \end{itemize}
\end{itemize}
where $\data(t) = \left\lbrace \obs(1),\ldots,\obs(t),\vect{\forcing}(1),\ldots,\vect{\forcing}(t) \right\rbrace$ is the available data up to time $t$, $\obs(t)$ ($t=1,\ldots,\tau_H$) are the observations, and $\vect{\forcing}(t) = \left[ \forcing_C(t), \forcing_V(t), \shared^\star(t) \right]^\prime$ is the forcing vector.
The shared discrepancy $\shared(t)$ enters the model with the other common components of forcing.
The volcanic forcing $\forcing_V(t)$ is set to zero for the future period.
By repeating the sampling procedure we can obtain as many samples $\obs^\star(\tau_H+1),\ldots,\obs^\star(\tau_F)$ of the future climate as we require.

\subsection{Discussion}
\label{sec:discussion}

When learning about the real-world, we only have 170 years of observations from the historical period to learn both the basic EBM parameters and the volcanic forcing and independent discrepancy parameters $F_V$ and $\sigma_\disc$.
Therefore, using the models to estimate the prior for the parameters $\latent_\obs$ in Equation~\ref{eqn:obs_par} is critical for obtaining realistic inferences since the observations will provide only limited information.
The fact that the shared discrepancy $\shared(t)$ is assumed to also apply to the real-world reflects the fact that $\shared(t)$ quantifies inadequacy in the ability of the EBM representations to approximate the more complete climate models.
Any inadequacy in the climate models ability to approximate the real-world is accounted for by the inflation of the prior on the real-world EBM parameters $\latent_\obs$ in Equation~\ref{eqn:obs_par}.

The initial conditions for the EBM representations of the climate model outputs are well defined since the models are all initialised from a 500 year run under pre-industrial conditions after a lengthy spin-up which should ensure they are (almost) in equilibrium.
However, for the real-world we have no observations for the pre-industrial period so we are forced to use the early industrial period 1850--1900 as a reference.
Also, we cannot be certain that the real system was in equilibrium prior to 1850.
The probabilistic representation used here means that the sensitivity of the projections to these assumptions can be explored through careful specification of the priors for the initial state, although we do not do so here.

The forcing discrepancy specified in Equations~\ref{eqn:disc} and \ref{eqn:shared} is among the simplest formulations that could address the biases seen in Figure~\ref{fig:reliability}.
Since the shared discrepancy $\shared(t)$ is interpreted as a common component of forcing, then arguably we could include model-specific radiative coefficients $F_\shared$, similar to $F_C$ and $F_V$.
However, this adds an additional parameter for each model (and the real-world) to an already over-parameterised system, when the model-specific discrepancy variances $\sigma_\disc^2$ already allow the scale of each model's response to vary uniquely.

The formulation in terms of random walks imparts little prior information except a degree of smoothness controlled by the variances $\sigma_\disc^2$ and $\sigma_\shared^2$.
Since the shared discrepancy $\shared(t)$ itself, not just its variance $\sigma_\shared^2$, is learned from the models and then used to drive the projections of the real-world, the formulation of $\shared(t)$ has little effect on the projections provided it is sufficiently general to capture the underlying behaviour.
However, the projections will be more sensitive to the formulation of the model and real-world specific discrepancies $\disc(t)$.
The random walk formulation implies that uncertainty about the state of the system will continue to increase even after the system has reached a new equilibrium.
For very long range projections, e.g., several centuries, this behaviour is obviously undesirable.
However, cross-validation indicates that the linear growth in uncertainty over time implied by this formulation is realistic for the RCP4.5 scenario as far as the year 2100.
If longer range projections are required, or the random walk formulation does not suit a particular scenario, then a more complex discrepancy formulation with time-varying variance could be fitted to the climate model output.
This would allow for additional uncertainty surrounding the timing of certain events, e.g., ice sheet melting, or a reduction in uncertainty once a new equilibrium was reached.

The assumption of exchangeability between the CMIP5 models in Section~\ref{sec:ensemble} implies that we would also specify the same prior beliefs about every pair, triple, etc. of models.
In practical terms, this means that each model should be equally similar to every other model.
For climate models this is clearly not the case. 
Some centres submit more than one model, or more than one version of the same model, some models from different centres share whole atmosphere or ocean component models.
These models will be more similar than those that share no common components.
To satisfy the assumption of exchangeability we analyse only a subset of the available models that we judge to be approximately exchangeable.

\section{Data, inference and model checking}
\label{sec:data-inf}

\subsection{Data}
\label{sec:data}

For each CMIP5 model we select run \textit{r1i1p1} from the pre-industrial control (piControl), \abrupt\ (abrupt4xCO2), historical and RCP4.5 (rcp45) experiments.
The variables used are near surface temperature ($tas$) and the top-atmosphere radiation balance ($rsdt - rsut - rlut$).
Data are globally and annually averaged to give bivariate time series of 500~years for the piControl experiment, 150~years for the \abrupt\ experiment, and 251~years for the combined historical and RCP4.5 experiments.
Some models have missing years at the end of the \abrupt\ experiment or the beginning of the historical experiment.
The Kalman filter methodology used to fit the EBMs (see Appendix~\ref{app:disc}) can handle these missing values without special provision.
In order to fit EBM representations to the model output we require temperature and radiation anomalies relative to an equilibrium state.
Therefore, the mean of the piControl for each model is removed from the outputs of the \abrupt, historical and RCP4.5 experiments.


The assumption of exchangeability between models in Section~\ref{sec:ensemble} implies that every model should be equally similar to every other model.
In order to satisfy this assumption, we analyse only a subset of the available models.
The 13 models chosen are listed in Table~\ref{tab:models}.
The subset is based on the thinned ensemble analysed by \citet{Sansometal2020}, where the models were chosen to minimise common components between models while maintaining similar horizontal and vertical resolutions.
There are three differences compared to the ensemble analysed by \citet{Sansometal2020}.
The CCSM4 model based on the older CAM4 atmosphere models has been substituted for the more recent CESM1 model based on the updated CAM5 atmosphere, since not all the required runs were available from the CESM1 model.
Similarly, the EC-EARTH model is missing due to a missing file in one of the required runs.
Finally, the INM-CM4 model was excluded since it did not include volcanic forcing in the historical experiment.
The chosen models are also similar to those analysed by \citet{Cumminsetal2020} with the exception that we include GFDL-ESM2G and IPSL-CM5A-MR rather than GFDL-ESM2M and IPSL-CM5A-LR respectively.

\begin{table}[t]
 \caption{CMIP5 models.}
  \begin{tabular}{lll}
  \hline
  Centre & Model & Institution \\
  \hline
  BCC   & BCC-CSM1.1   & Beijing Climate Center, China \\
  CCCma & CanESM2      & Canadian Centre for Climate Modelling and Analysis, Canada \\
  NCAR  & CCSM4        & National Center for Atmospheric Research (NCAR), United States \\
  CNRM  & CNRM-CM5     & Centre National de Recherches M\`{e}t\`{e}orologiques, France \\
  LASG  & FGOALS-s2    & Institute of Atmospheric Physics, China \\
  GFDL  & GFDL-ESM2G   & Geophysical Fluid Dynamics Laboratory, United States\\
  GISS  & GISS-E2-R    & NASA Goddard Institute for Space Studies, United States\\
  MOHC  & HadGEM2-ES   & Met Office Hadley Centre, United Kingdom \\
  IPSL  & IPSL-CM5A-MR & Institut Pierre-Simon Laplace, France \\
  MIROC & MIROC5       & Japan Agency for Marine-Earth Science and Technology, Japan \\ 
  MPI-M & MPI-ESM-LR   & Max Planck Institute for Meteorology, Germany \\
  MRI   & MRI-CGCM3    & Meteorological Research Institute, Japan\\
  NCC   & NorESM1-M    & Norwegian Climate Centre, Norway \\
\hline
  \end{tabular}
  \label{tab:models}
\end{table}

For the real-world, we use the annual ensemble median GMST from the HadCRUT4 dataset \citep{Moriceetal2012}.
The HadCRUT4 dataset provides anomalies relative to the 1961--1990 average, so the anomalies need to be adjusted for compatibility with the models.
Since observations prior to 1850 are not readily available, we adopt the IPCC SR1.5 approach and re-reference the anomalies to the 1850--1900 average \citep{IPCC2018}.
HadCRUT4 also includes an extensive quantification of the uncertainties associated with the observations.
We use the provided lower and upper 95\,\% confidence bounds for the combined effects of all of the uncertainties on the annual time series to compute the annual standard deviation of the observation uncertainty assuming the observations follow a normal distribution.
We use these standard deviations as our estimate of the independent annual observation uncertainty.
This is likely to be an over-estimate of the implied uncertainty since the true uncertainty is likely to be correlated between years.
Although observations of top-of-atmosphere radiation do exist, they are limited to the satellite era (1970s onwards) making them difficult to use in this context, so we choose not to include them.

To drive the EBM representations we use the \cotwo\ equivalence concentrations from the CMIP5 concentrations datasets \citep{Meinshausenetal2011}.
Although most major forcings were specified for the CMIP5 experiments, stratospheric injection of sulfate aerosols from explosive volcanic eruptions was not \citep{Driscolletal2012}.
However, most modelling groups chose to impose the stratospheric emissions from volcanic eruptions and the effects are clearly visible in Figure~\ref{fig:reliability}.
To account for stratospheric aerosol emissions, we use the updated global mean stratospheric aerosol optical depth at \SI{550}{\nano\meter} dataset by \citet{Satoetal1993} available from NASA GISS (\url{https://data.giss.nasa.gov/modelforce/strataer/}).

\subsection{Inference}
\label{sec:inference}

Our aim is to evaluate $\Pr \left( \obs_F \mid \forcing_F \right)$, the distribution of future climate given $\obs_F = \lbrace \obs(\tau_H+1),\ldots,\obs(\tau_F) \rbrace$ given future \cotwo\ forcing $\forcing_F = \lbrace \forcing_C(\tau_H+1),\ldots,\forcing_C(\tau_F) \rbrace$.
In Section~\ref{sec:projection}, we outlined how this can be done by Monte Carlo methods, sampling from
\begin{align}
  \Pr \left( \obs_F \mid \forcing_F \right) 
    = \int 
      \Pr \left[ \obs_F \mid 
        \latent_\obs, \shared_F, \state_\obs(\tau_H), \forcing_F \right] 
      \dd \latent_\obs, \shared_F, \state_\obs,
\end{align}
where $\latent_\obs$ are the parameters of the EBM representation of the real-world, $\shared_F = \lbrace \shared(\tau_H+1),\ldots,\shared(\tau_F) \rbrace$ is the shared discrepancy, and $\state_\obs(\tau_H) = \left[ F(\tau_H),T_1(\tau_H),T_2(\tau_H),T_3(\tau_H),\disc(\tau_H) \right]^\prime$ is the state of the real-world at the end of the historical/observed period.

However, we still need to be able to evaluate the posterior distribution of the EBM parameters $\latent_\obs$ and the shared discrepancy $\shared$.
This is achieved by Markov-Chain Monte Carlo sampling from the full posterior
\begin{align*}
  \Pr \left( \latent_\obs, \shared \mid \data \right) 
    = \int \Pr \left( \latent_\obs, \shared, \latent_1,\ldots,\latent_M, \vect{\mu}_\latent, \mat{\Sigma}_\latent, \sigma_\shared \mid Z_H, \kappa, \data \right) \dd \latent_1,\ldots,\latent_M, \vect{\mu}_\latent, \mat{\Sigma}_\latent, \sigma_\shared
\end{align*}
where $\data$ represents the available data, i.e., model outputs from the \abrupt, historical and RCP4.5 experiments, and \cotwo\ and volcanic aerosol forcings for those experiments.
In practice, we assume that
\begin{multline}
 \Pr \left( \latent_\obs, \shared, \latent_1,\ldots,\latent_M, \vect{\mu}_\latent, \mat{\Sigma}_\latent, \sigma_\shared \mid Z_H, \kappa, \data \right) 
   = \\ 
   \Pr \left( \latent_\obs \mid \shared, \vect{\mu}_\latent, \mat{\Sigma}_\latent, \kappa, Z_H, X_H \right)
   \Pr \left( \shared, \latent_1,\ldots,\latent_M, \vect{\mu}_\latent, \mat{\Sigma}_\latent, \sigma_\shared \mid \data \right).
\end{multline}
This implies that the observations $Z_H$ do not contribute to the estimation of the shared discrepancy $\shared$, its variance $\sigma_\shared^2$, or the common parameters $\vect{\mu}_\latent$ and $\mat{\Sigma}_\latent$.
This assumption is not necessary, but is stated for transparency, and intended to emphasise the role of the models in providing prior information for inference about the real-world.
Full details of the priors for $\vect{\mu}_\latent$, $\mat{\Sigma}_\latent$ and $\sigma_\shared$, and the partially collapsed Gibb's sampler used to sample the full posterior are given in the Supplementary Material.
The process is simplified by conditioning on the shared discrepancy $\shared$, as we do when sampling the future climate in Section~\ref{sec:projection}.
Due to the inclusion of $\shared$, Equations~\ref{eqn:forcing}--\ref{eqn:radiation} imply a $5M$ dimensional state-space model for the \emph{joint} distribution of the outputs from the climate models.
The resulting joint likelihood function for $\latent_1,\ldots,\latent_M$ would be difficult and expensive to evaluate.
Conditioning on $\shared$ simplifies the process by allowing us to fit $M$ independent 5 dimensional state-space models (see Appendix~\ref{app:full}) and evaluate $M$ much simpler likelihood functions instead, one for each climate model.

The results in Section~\ref{sec:results} are based on samples from four parallel chains, initialised from well dispersed starting conditions (see Supplementary Material for details of initialisation).
Each chain was run for a burn-in period of 25\,000 samples during which robust adaptive Metropolis-Hastings sampling \citep{Vihola2012} was used to learn optimal proposal distributions for $\latent_1,\ldots,\latent_M$ and $\sigma_\shared$.
Each chain was then run for another 250\,000 samples with the proposal distributions fixed at their values at the end of the burn-in period.
Gelman-Rubin diagnostics \citep{GelmanRubin1992} and visual inspection indicated that all variables converged successfully.
The eight chains give a total of 2\,000\,000 samples of each variable, however for final storage only every 200th sample was kept.
Therefore, the results in Section~\ref{sec:results} are based on 10\,000 samples and 10\,000 corresponding future trajectories.
All results are based on $\kappa = 1.0$ in Equation~\ref{eqn:obs_par} unless stated otherwise, i.e., the real-world is exchangeable with the models.

\subsection{Model checking}
\label{sec:checking}

In order to check whether the addition of the forcing discrepancy produces reliable estimates of future climate we performed a leave-one-out cross-validation.
When trying to project the climate of the real-world, only observations of surface temperature for the historical period are available.
Therefore, each CMIP5 model was excluded in turn, and only its output for surface temperature during the historical period was used in place of observations $\obs_H$ of the real-world.
The same MCMC procedure was then used to sample the posterior distribution of the parameters, but with only 4 chains and shorter burn-in and sampling periods of 25\,000 and 100\,000 samples respectively.
For each excluded model, we retain every 40th sample to give 10\,000 samples in total, and project forward using the methodology described in Section~\ref{sec:projection} to give 10\,000 future trajectories.
Standardised projection errors are computed as described in Appendix~\ref{app:proj}.

\section{Results}
\label{sec:results}

The posterior means of the EBM parameters for each climate model, and those for the real-world are shown in Table~\ref{tab:parameters}.
Overall, the posterior mean estimates of the main EBM parameters (excluding $F_V$ and $\sigma_\disc$) are very similar to those of \citet{Cumminsetal2020}.
This emphasises that most of the information about the EBM parameters still comes from the \abrupt\ experiment.
There were two unusual parameter estimates reported by \citet{Cumminsetal2020}, CNRM-CM5.1 had an unusually large value for $\gamma$ relative to the other models, and GISS-E2-R reported a very high value for $k_3$.
The regularisation imposed by the common prior over the models in Equation~\ref{eqn:params} has constrained both of these parameters to values more similar to the other models, although the value of $k_3$ for GISS-E2-R is still almost double the next closest model.

\begin{table}[t]
  \caption{Parameter estimates. The posterior means of the individual model parameters, the CMIP5 ensemble parameters and the observation parameters.}
  \begin{tabular}{lcccccccccccccc}
  \hline
  Model & $\gamma$ & $C_1$ & $C_2$ & $C_3$ & $k_1$ & $k_2$ & $k_3$ & $\varepsilon$ & $\sigma_F$ & $\sigma_T$ & $F_C$ & $F_V$ & $\sigma_\disc$ \\
  \hline
BCC-CSM1.1   & 3.50 & 3.93 &  9.4 &  53 & 1.23 & 2.71 & 0.69 & 1.28 & 0.64 & 0.37 & 3.60 & 23.8 & 0.048 \\
CanESM2      & 2.03 & 4.02 & 11.2 &  71 & 0.98 & 2.12 & 0.74 & 1.32 & 0.61 & 0.54 & 3.94 & 18.6 & 0.035 \\
CCSM4        & 2.52 & 4.40 & 13.4 &  78 & 1.29 & 2.03 & 1.22 & 1.37 & 0.62 & 0.51 & 4.06 & 23.1 & 0.039 \\
CNRM-CM5     & 3.54 & 3.55 & 10.2 &  82 & 1.14 & 2.66 & 0.64 & 0.97 & 0.56 & 0.44 & 3.66 & 21.0 & 0.048 \\
FGOALS-s2    & 2.05 & 4.56 & 11.5 & 134 & 0.83 & 1.67 & 1.17 & 1.28 & 0.72 & 0.64 & 3.90 & 20.0 & 0.034 \\
GFDL-ESM2G   & 2.36 & 4.74 & 14.8 & 104 & 1.46 & 1.79 & 1.48 & 1.31 & 0.71 & 0.54 & 3.63 & 23.1 & 0.038 \\
GISS-E2-R    & 2.80 & 5.09 & 29.0 & 118 & 1.78 & 1.93 & 3.31 & 1.44 & 0.45 & 0.35 & 4.11 & 23.4 & 0.037 \\
HadGEM2-ES   & 1.96 & 4.13 & 10.0 &  91 & 0.62 & 2.34 & 0.66 & 1.37 & 0.60 & 0.40 & 3.20 & 14.8 & 0.031 \\
IPSL-CM5A-MR & 2.27 & 3.90 & 12.0 &  95 & 0.79 & 2.45 & 0.75 & 1.19 & 0.52 & 0.41 & 3.47 & 16.1 & 0.034 \\
MIROC5       & 1.58 & 4.43 & 22.6 & 130 & 1.61 & 1.55 & 1.77 & 1.17 & 0.52 & 0.80 & 4.45 & 19.3 & 0.032 \\
MPI-ESM-LR   & 2.05 & 4.08 & 13.1 &  74 & 1.13 & 1.98 & 0.95 & 1.33 & 0.57 & 0.61 & 4.38 & 20.3 & 0.036 \\
MRI-CGCM3    & 2.18 & 3.96 & 13.5 &  66 & 1.20 & 2.62 & 0.74 & 1.27 & 0.54 & 0.39 & 3.32 & 16.6 & 0.036 \\
NorESM1-M    & 1.85 & 4.74 & 17.2 & 107 & 1.11 & 2.05 & 1.44 & 1.42 & 0.57 & 0.44 & 3.45 & 16.7 & 0.030 \\
\hline
Ensemble     & 2.37 & 4.29 & 14.5 &  94 & 1.18 & 2.16 & 1.21 & 1.29 & 0.59 & 0.50 & 3.78 & 19.8 & 0.037 \\
\hline
Observations & 2.07 & 4.39 & 17.0 & 102 & 1.17 & 2.24 & 1.32 & 1.34 & 0.54 & 0.45 & 3.59 & 17.2 & 0.033 \\
\hline
  \end{tabular}
  \label{tab:parameters}
\end{table}

\subsection{Model checking}
\label{sec:cv}

\begin{figure}[t]
  \centering\includegraphics[width=0.5\textwidth]{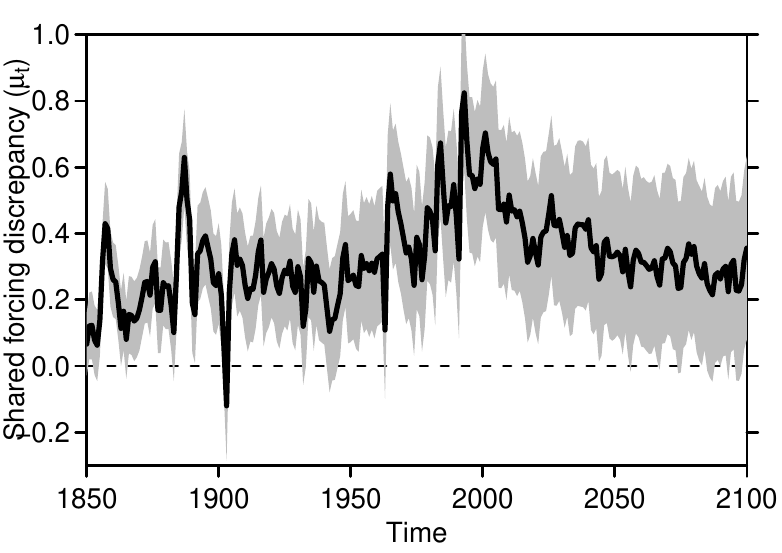}
  \caption{The shared forcing discrepancy. The posterior mean (solid line) and marginal 90\,\% credible intervals (shading) for the shared forcing discrepancy $\mu_t$ under the RCP4.5 forcing scenario.}
  \label{fig:shared}
\end{figure}

\begin{figure}[t]
  \centering\includegraphics[width=\textwidth]{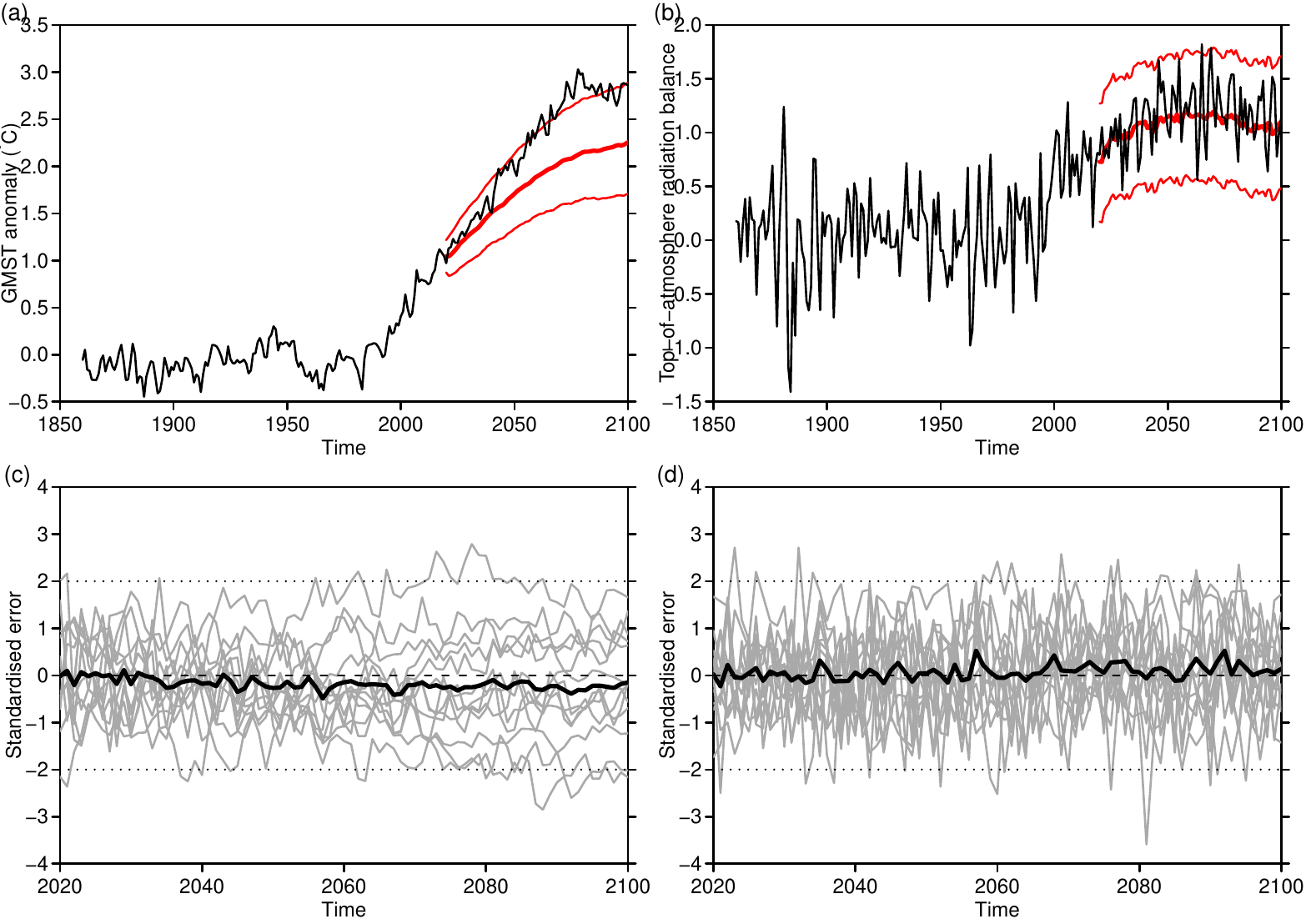}
  \caption{Cross validation.
           (a) and (b) global mean surface temperature and top-of-atmosphere radiation balance respectively in the HadGEM2-ES model under the RCP4.5 scenario.
           Black lines are the model simulations, thick red lines are the posterior predictive means based on the model including both individual and shared forcing discrepancies, and thin red lines are a marginal 90\,\% credible interval.
           (c) and (d) Standardised prediction errors for global mean surface temperature and top-of-atmosphere radiation balance respectively under the RCP4.5 scenario including both individual and shared forcing discrepancies.
           Thin grey lines represent standardised errors from individual CMIP5 models.
           The thick black line is the ensemble mean.}
  \label{fig:cv}
\end{figure}

The posterior distribution of the shared discrepancy $\mu(t)$ is plotted in Figure~\ref{fig:shared} and roughly follows the pattern seen in Figure~\ref{fig:reliability}, peaking around the year 2000 before slowly declining.
The 10\,000 future trajectories from each model sample the posterior predictive distribution for that model.
An example of the posterior predictive distributions is shown for HadGEM2-ES in Figure~\ref{fig:cv}(a) and (b).
The temperature projections are biased low, but the credible interval now includes most of the model output, and the radiation balance is still well predicted.

To check the overall reliability of the projections we plot the standardised predictive error of each model from the cross-validation at times $t=2020,\ldots,2100$ in Figure~\ref{fig:cv}(c) and (d).
If the projections are reliable, then approximately 95\,\% of the standardised errors should lie between $-2$ and $+2$ standard deviations at all times.
For surface temperature, we see this is approximately the case.
The projections appear very reliable, with good scatter between $-2$ and $+2$ standard errors and almost no overall mean bias.
This contrasts sharply with Figure~\ref{fig:reliability}(c), indicating much improved projections.
For the radiation balance in Figure~\ref{fig:cv}(d), the projections also appear reliable, with little or no mean bias.

The cross-validation gives us confidence that even by assimilating only observations of temperature for the real-world we can obtain reliable projections of future surface temperature.
The mean bias has been almost completely removed and the projections now appear reliable or slightly under-confident rather than very over-confident as in Figure~\ref{fig:reliability}.

The Supplementary Material also includes extensive additional analysis checking the sensitivity of our inferences to our choice of priors, choice of climate models and choice of the coexchangeable coefficient $\kappa$.
Our inferences are insensitive to the choice of priors and surprisingly insensitive to the coexchangeable coefficient $\kappa$.
Inferences are not strongly influenced by the choice of climate models, but the sensitivity analysis does highlight the potential for bias due to including multiple variants of the same model.

\subsection{Equilibrium Climate Sensitivity}
\label{sec:ecs}

\begin{figure}[t]
  \centering\includegraphics[width=0.5\textwidth]{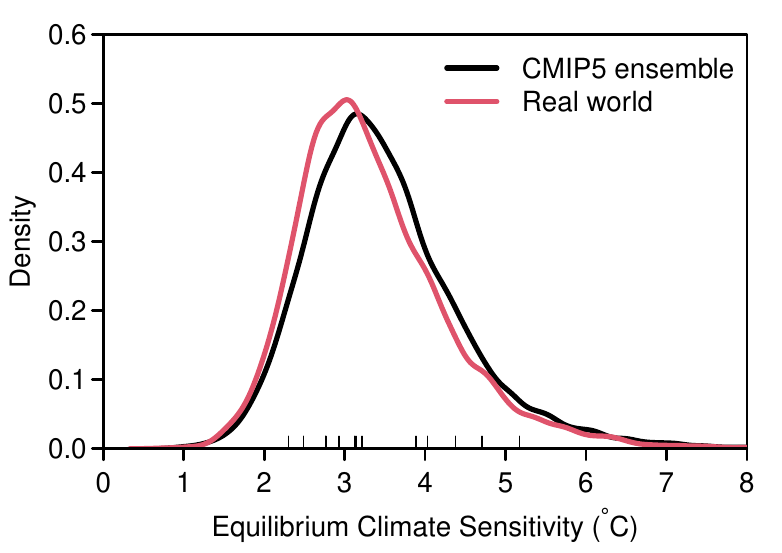}
  \caption{Equilibrium climate sensitivity. The posterior density of the equilibrium climate sensitivity for the real-world (red) and the CMIP5 ensemble (black).}
  \label{fig:ecs}
\end{figure}

The posterior mean estimates of the parameters of the real-world in Table~\ref{tab:parameters} differ very little from those of the ensemble given by $\params$ with the exception of the heat transfer coefficients $k_1$, $k_2$ and $k_3$.
Figure~\ref{fig:ecs} shows the posterior distribution of the ECS of the real-world, given by $F_C / k_1$.
The posterior distribution of ECS for the CMIP5 ensemble is also shown, estimated by sampling new values of $F_C$ and $k_1$ from Equation~\ref{eqn:params} conditional on the posterior samples of $\params$.
Due to the limited signal in the historical temperature observations, there is insufficient information to usefully constrain the ECS of the real world compared to the CMIP5 ensemble.
For the real-world, we estimate a median ECS of 3.2\,K and 90\,\% credible interval 2.1--5.1\,K.
For the CMIP5 ensemble we estimate a median ECS of 3.3\,K and 90\,\% credible interval 2.1--5.3\,K.

Various authors have tried to constrain estimates of ECS using a variety of metrics, see \citet{Brient2020} or \citet{Halletal2019} and references therein for examples, although the credibility of some of these estimates has been questioned \citep{Caldwell2018}.
Table~\ref{tab:ecs} compares our estimate with that of several recent studies, including the synthesis report by \citet{Sherwood2020}.
Compared to the previous Bayesian hierarchical analysis by \citet{Jonkoetal2018}, our median estimate is higher, although our credible interval similar in width.
This study was not targeted specifically at constraining ECS, so it is not surprising that other studies have proposed estimates that differ more strongly from the median of the models.
However, both our median estimate and credible interval are very similar to those of the synthesis report by \citet{Sherwood2020}.

\begin{table}[t]
  \caption{Estimates of Equilibrium Climate Sensitivity from the CMIP5 ensemble. Note that the IPCC interval is the ``likely'' 66\,\% interval, not a 90\,\% interval.}
  \begin{tabular}{lll}
  \hline
  Study & Median & \SIrange{5}{95}{\percent} \\
  \hline
  IPCC & & \SIrange{1.5}{4.5}{\kelvin} \\
  Cox et al.~\cite{Cox2018} & \SI{2.8}{\kelvin} & \SIrange{1.6}{4.0}{\kelvin} \\
  Jonko et al.\cite{Jonkoetal2018} & \SI{2.5}{\kelvin} & \SIrange{1.2}{3.9}{\kelvin} \\
  Jiménez-de-la-Cuesta \& Mauritsen~\cite{Jimenez2019} & \SI{2.8}{\kelvin} & \SIrange{1.7}{4.1}{\kelvin} \\
  Nijsse et al.~\cite{Nijsse2020} & \SI{2.3}{\kelvin} & \SIrange{1.0}{4.1}{\kelvin} \\
  Sherwood et al.~\cite{Sherwood2020} & \SI{3.1}{\kelvin} & \SIrange{2.3}{4.7}{\kelvin} \\
  This study & \SI{3.2}{\kelvin} & \SIrange{2.1}{5.1}{\kelvin} \\
  \hline
  \end{tabular}
  \label{tab:ecs}
\end{table}


\subsection{Future projections}
\label{sec:projections}

\begin{figure}[t]
  \centering\includegraphics[width=0.5\textwidth]{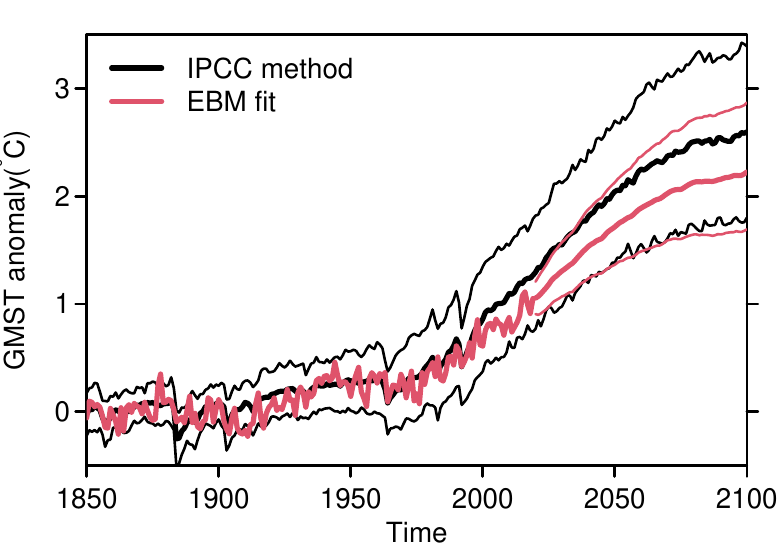}
  \caption{Global mean surface temperature anomaly projections. 
  Posterior predictive distribution for global mean surface temperature under  the RCP4.5 forcing scenario. 
  Thick lines indicate the posterior predictive means of the CMIP5 ensemble (black) and the real-world (red). 
  Thin lines indicate marginal 90\,\% credible intervals. Grey lines are the individual CMIP5 model simulations.}
  \label{fig:projections}
\end{figure}

Figure~\ref{fig:projections} shows the projections for the real-world under the RCP4.5 scenario based on the EBM fit to historical observations and accounting for shared and unique forcing bias.
The projections make the usual assumption that the real-world is exchangeable with the climate model ensemble, i.e., $\kappa = 1.0$.
The projections are well constrained and lie entirely in the lower half of the range predicted by the CMIP5 ensemble.
The mean surface temperature increase above pre-industrial conditions projected in 2100 is \SI{2.2}{\kelvin} with 90\,\% credible interval \SIrange{1.7}{2.9}{\kelvin}.
This compares with an enlarged CMIP5 ensemble (sampling new models from Equation~\ref{eqn:params}) with mean \SI{2.5}{\kelvin} with 90\,\% credible interval \SIrange{1.6}{4.0}{\kelvin} in 2100, and an IPCC-method estimate of \SI{2.6}{\kelvin} with 90\,\% credible interval \SIrange{1.8}{3.4}{\kelvin}.


Table~\ref{tab:rcp45} compares our projections of the mean warming in the period 2081-2100 above the 1986-2005 average with other recent studies.
Our estimated median warming of \SI{1.5}{\kelvin} is \SI{0.3}{\kelvin} lower than the other recent estimates which all agree on warming of around \SI{1.8}{\kelvin} by the end of the century.
Our credible interval is also narrower than those of the IPCC or \citet{Tokarska2020}, similar to the synthesis estimate of \citet{Sherwood2020}, but wider than that of \citet{Strobach2020}.

\begin{table}[t]
  \caption{Estimates of future warming under RCP4.5 scenario averaged over 2081--2100. Note that the IPCC and Sherwood et al.~\cite{Sherwood2020} intervals are 66\,\% intervals, not 90\,\% intervals.}
  \begin{tabular}{lllll}
  \hline
  Study & Ensemble & Reference & Median & \SIrange{5}{95}{\percent} \\
  \hline
  IPCC & CMIP5 & 1986--2005 & \SI{1.8}{\kelvin} & \SIrange{1.1}{2.6}{\kelvin} \\
  Sherwood et al.~\cite{Sherwood2020} & CMIP5 & 1986--2005 & \SI{1.8}{\kelvin} & \SIrange{1.4}{2.3}{\kelvin} \\
  Sherwood et al.~\cite{Sherwood2020} & CMIP5 & 1986--2005 & \SI{1.8}{\kelvin} & \SIrange{1.7}{2.1}{\kelvin} \\
  Tokarska et al.~\cite{Tokarska2020} & CMIP6 & 1995--2014 & \SI{1.8}{\kelvin} & \SIrange{1.2}{2.5}{\kelvin} \\
  This study & CMIP5 & 1986--2005 & \SI{1.5}{\kelvin} & \SIrange{1.1}{2.1}{\kelvin} \\
  \hline
  \end{tabular}
  \label{tab:rcp45}
\end{table}

\citet{Jonkoetal2018} only made projections under the stronger RCP8.5 scenario.
Our full analysis of the RCP8.5 scenario is included in the Supplementary Material.
Under the RCP8.5 forcing scenario, \citet{Jonkoetal2018} estimate a 90\,\% credible interval of 2.2\,K--5.6\,K in the year 2100 compared to the pre-industrial period (no median estimate was given).
In comparison, we estimate a median warming of 4.3\,K with 90\,\% credible interval 3.4\,K--5.6\,K.
Given the positive skewness in the majority of estimates, our median warming is likely higher than that of \citet{Jonkoetal2018}, but our credible interval is much narrower, despite the methodological similarity.
The difference in the median can be explained by our inclusion of the shared forcing discrepancy, correcting for the tendency of the EBMs to underestimate the warming in Figure~\ref{fig:reliability}.
The difference in credible interval is likely due to the fact that \citet{Jonkoetal2018} estimate the distribution over the models in Equation~\ref{eqn:params} from the distribution of the individual EBM parameter estimates, whereas we learn both the individual and ensemble parameters simultaneously.
This results in some regularisation (shrinkage) of the individual estimates towards the consensus of the ensemble, and so a tighter distribution over the models.
Since the model distribution acts as a prior for the real-world, this in turn results in a tighter distribution for the real-world. 

\begin{figure}[t]
  \centering\includegraphics[width=\textwidth]{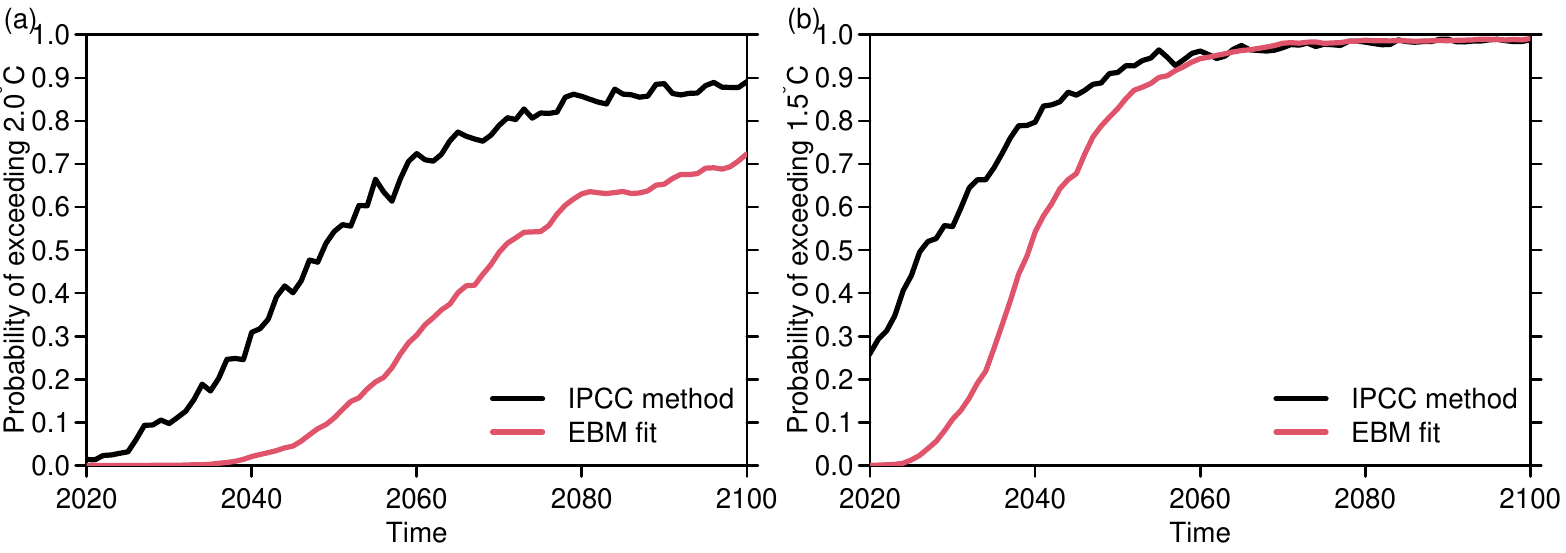}
  \caption{Probability of meeting the Paris agreement. 
 (a) The probability of climate change below $2.0\,^\circ\mathrm{C}$ under the RCP4.5 forcing scenario, and
 (b) the probability of climate change below $1.5\,^\circ\mathrm{C}$ under the RCP4.5 forcing scenario.}
  \label{fig:probabilities}
\end{figure}

Figure~\ref{fig:probabilities} shows the probabilities of meeting the targets set out in the Paris Agreement under the RCP4.5 scenario.
Under the constrained projections, there is almost no probability of GMST exceeding \SI{2.0}{\kelvin} before 2040, after which the probability rises rapidly at first then more slowly until it reaches 0.72 in 2100 in Figure~\ref{fig:probabilities}(a).
This is in striking contrast to the IPCC anomaly method which gives a probability of 0.89.
The outlook for remaining below \SI{1.5}{\kelvin} is less optimistic.
The constrained projections agree with the anomaly projections that under the RCP4.5 scenario there is a probability of 0.99 that GMST will exceed \SI{1.5}{\kelvin} above pre-industrial levels in 2100.

\section{Conclusions}
\label{sec:conclusions}

In this study we propose a Bayesian hierarchical approach to projecting future GMST by combining outputs from multiple climate models and observations of the real-world.
Our approach builds on existing methods for combining projections from multiple climate models by incorporating physically motivated representations of climate model outputs and observations.
An additional innovation is the inclusion of a discrepancy in the effective forcing due to processes not captured by the simplified representation but whose effects may vary between forcing scenarios leading to biased projections.
The proposed methodology not only provides point projections, but associated credible intervals and probabilities, accounting for natural variability, observation uncertainty, structural uncertainty and model inadequacy.

Compared to existing heuristic anomaly methods, observations of the real-world are integral to the projections, rather than simply an offset applied to an ensemble of models.
Observations of the historical period contain limited information about the EBM representation.
However, we have shown that by combining a suitable prior based on climate model outputs, there is enough information to usefully constrain projections of future climate.
Although our proposed method still relies on anomalies relative to a fixed reference period, that period is now motivated by physical concerns, i.e., approximate equilibrium.
Previously, the reference period would often be chosen to be as close to the present day as possible in order to limit the divergence of projections from different models.
In contrast, we move the reference period as far back in time as possible and assimilate a full 170 years of annual observations rather than a single 30-year mean.
Moving between reference periods for the sake of expressing results is now simply a matter of subtracting the appropriate offset between the means of the reference periods
Unlike commonly used heuristic methods, the uncertainty of the projections does not change.


In forming our projections, we only used one run of each scenario from each model, when several runs of some scenarios are available from some models.
By not using all available runs we are potentially throwing away valuable information.
The methodology proposed here could easily be expanded to include multiple initial conditions runs without biasing our inferences towards the models with the most runs.
However, the EBM representation arguably makes the inclusion of multiple runs of less value than might otherwise be the case.
The EBM representation is a linear dynamic system.
Therefore the response to a linear combination of inputs, i.e., known forcing and natural variability, is equal to the sum of the responses of the individual inputs.
So by learning the parametric representation of the forced response, we are simultaneously learning the response to natural variability and vice-versa. 

The problem of including all available models is more complicated.
In order to avoid biasing our projections towards models or components that are over-represented in the CMIP5 ensemble, we chose to infer our prior for the real-world based on only a subset of the available climate models.
In doing so, we risk losing valuable information contained in the excluded models.
However, comparing projections from the subset against projections using the full ensemble suggests that any information loss is very limited.

Our results demonstrate that by working with physically interpretable representations and parameters, it is possible to obtain strongly constrained projections without the need to adopt performance based model weights.
Having methods available that make very different assumptions is a healthy thing, since those assumptions can then be challenged and tested.
The strength of the model weighting approach is that all available models can be included in the analysis.
Further research is required to combine prior knowledge with dependences diagnosed from model outputs to enable the inclusion of all available models within formal statistical frameworks.

In principle, the methodology proposed here could be modified to project any future  \cotwo\ emissions scenario.
Due to the presence of shared biases in the effective forcing, we chose to learn a shared component of forcing from simulations of the future as well as historical period.
This makes our parameter inference and projections specific to a particular emissions scenario and limits us to making projections of scenarios for which we have model outputs to learn from.
However, if the shared discrepancy and model-specific discrepancy parameters were only learned from the historical period, the simulations could then be allowed to evolve freely following any future emissions scenario.
The resulting projections would have greater uncertainty than those shown here in order to account for future changes in the shared component which is then treated as unknown.
However, both the shared and specific discrepancies would require more careful specification to ensure that projections are both credible and reliable.
We have shown that reliable and well constrained century scale projections are possible using simulations from only a few climate models, without having to make these additional assumptions.
Further research is required to holistically quantify the effects of uncertainty about future emissions.


\section*{Acknowledgement}

We acknowledge the World Climate Research Programme's Working Group on Coupled Modelling, which is responsible for CMIP, and we thank the climate modeling groups (listed in Table~\ref{tab:models} of this paper) for producing and making available their model output. For CMIP the U.S. Department of Energy's Program for Climate Model Diagnosis and Intercomparison provides coordinating support and led development of software infrastructure in partnership with the Global Organization for Earth System Science Portals.

\appendix

\section*{Appendix}

\section{State-space representation and discretisation}
\label{app:disc}

Equations~\ref{eqn:t10}--\ref{eqn:forcing0} can be written as
\begin{align*}
  \frac{\dd \state}{\dd t}& = \evolution \state(t) + \evolutionreg \vect{\forcing}(t) + \vect{w}(t) & \vect{w}(t) & \sim \normal \left( \vect{0}, \evolutionvar \right)
\end{align*}
where
\begin{align*}
  \evolutionreg & = 
  \begin{bmatrix}
   \gamma F_C \\
   0 \\
   0 \\
   0 \\
  \end{bmatrix} &
  \vect{\forcing}(t) = 
  \begin{bmatrix}
    \forcing_C(t)
  \end{bmatrix}  
\end{align*}
and
\begin{align*}
  \evolution & =
  \begin{bmatrix}
    -\gamma & 0 & 0 & 0 \\
    \frac{1}{C_1} & -\frac{k_1 + k_2}{C_1} & \frac{k_2}{C_1} & 0 \\
    0 & \frac{k_2}{C_2} & -\frac{k_2 + \varepsilon k_3}{C_2} &  \frac{\varepsilon k_3}{C_2} \\
    0 & 0 & \frac{k_3}{C_3} & -\frac{k_3}{C_3} \\
  \end{bmatrix} &
  \evolutionvar & =
  \begin{bmatrix}
    \sigma_F^2 & 0 & 0 & 0 \\
    0 & \frac{\sigma_T^2}{C_1^2} & 0 & 0 \\
    0 & 0 & 0 & 0 \\
    0 & 0 & 0 & 0 \\
  \end{bmatrix}.
\end{align*}

Following \citet{Cumminsetal2020}, the discretised state space form of the EBM in Equations~\ref{eqn:forecast_eqn} and \ref{eqn:evolution_eqn} is
\begin{align*}
  \forecast_d &= 
  \begin{bmatrix}
    0 & 1 & 0 & 0 \\
    1 & -k_1 & (1 - \varepsilon) k_3 & -(1 - \varepsilon) k_3
  \end{bmatrix} &
  \forecastvar_d &= 
  \begin{bmatrix}
    0 & 0 \\
    0 & 0
  \end{bmatrix}
\end{align*}
where
\begin{align*}
  \evolution_d    & = 
    e^\evolution &
  \evolutionreg_d & = 
    \evolution^{-1} (\evolution_d - \mat{I}) \evolutionreg &
  \evolutionvar_d & = 
    \int_{s=0}^1 e^{\evolution s} \evolutionvar e^{\evolution^\prime s}
\end{align*}
are the discretised forms of the matrices $\evolution$, $\evolutionreg$ and $\evolutionvar$.

\subsection*{Model likelihood}

For parameter fitting by maximum likelihood in discrete time, the EBM likelihood factorises as
\begin{equation}
  \Pr \left[ \vect{\out}(1),\ldots,\vect{\out}(T) \mid \vect{\forcing}(1),\ldots,\vect{\forcing}(T), \latent \right]
    = \prod_{t=1}^T \Pr \left[ \vect{\out}(t) \mid \data(t-1), \latent \right]
  \label{eqn:likelihood}
\end{equation}
where $\data(t) = \left\lbrace \vect{\out}(1),\ldots,\vect{\out}(t), \vect{\forcing}(1),\ldots,\vect{\forcing}(t) \right\rbrace$ represents the data up to time $t$, and $\latent = ( \gamma, C_1, C_2, C_3, \allowbreak k_1, k_2, k_3, \varepsilon, \sigma_F, \sigma_T, F_C )^\prime$ is the vector of EBM parameters.
The likelihood can be efficiently evaluated by use of the Kalman filter.

\subsection*{The Kalman filter}
The Kalman filter was originally conceived to provide a best estimate of the state vector $\state(t)$ given all the data up to that time.
When formulated probabilistically from a Bayesian perspective, it provides prior and posterior distributions for the $\state(t)$ at each time $t$, and the prior distribution for the data $\vect{\out}(t)$ required to evaluate the likelihood.
The familiar prediction and update steps of the Kalman filter are then:

\subsubsection*{Prediction step}

The prior distribution for the state $\state$ at time $t$ given all previous data is
\begin{equation}
  \left[ \state(t) \mid \data(t-1), \latent \right] \sim \normal \left[ \priorexpstate(t), \priorvarstate(t) \right]
  \label{eqn:state_pred}
\end{equation}
where
\begin{align*}
  \priorexpstate(t) & = \evolution_d \postexpstate(t-1) + \evolutionreg_d \vect{\forcing}(t) &
  \priorvarstate(t) & = \evolution_d \postvarstate(t-1) \evolution_d^\prime + \evolutionvar_d
\end{align*}
and $\postexpstate(t-1)$ and $\postvarstate(t-1)$ are the posterior expectation and covariance of the state $\state$ at time $t-1$.

The prior distribution for the data $\vect{\out}$ at time $t$ is
\begin{equation}
  \left[ \vect{\out}(t) \mid \data(t-1), \latent \right] \sim \normal \left[ \priorexpobs(t), \priorvarobs(t) \right]
  \label{eqn:obs_pred}
\end{equation}
where
\begin{align*}
  \priorexpobs(t) & = \forecast_d \priorexpstate(t) & 
  \priorvarobs(t) & = \forecast_d \priorvarstate(t) \forecast_d^\prime + \forecastvar_d.
\end{align*}

\subsubsection*{Update step}

The posterior distribution of the state $\state$ at time $t$ given the data at time $t$ is then
\begin{equation}
  \left[ \state(t) \mid \data(t), \latent \right] \sim \normal \left[ \postexpstate(t), \postvarstate(t) \right]
  \label{eqn:state_post}
\end{equation}
where
\begin{align*}
  \postexpstate(t) & = \priorexpstate(t) + \kalmangain(t)  \left[ \vect{\out}(t) - \priorexpobs(t) \right] &
  \postvarstate(t) & = \priorvarstate(t) - \kalmangain(t) \priorvarobs(t) \kalmangain^\prime(t)
\end{align*}
and $\kalmangain(t) = \mat {R}(t) \forecast_d^\prime \mat{Q}(t)^{-1}$ is the Kalman gain at time $t$.

The expected state at time $t=0$ for the \abrupt\ experiment is given by $\postexpstate(0) = (2 F_C, 0 ,0 ,0)^\prime$, i.e., 4$\times$\cotwo\ forcing from Equation~\ref{eqn:co2} applied to an initial equilibrium state.
The covariance $\postvarstate(0)$ of the state at time $t= 0$ is taken to be the stationary marginal covariance of the EBM with only stochastic forcing \citep[Appendix C]{Cumminsetal2020}.

\section{Projections and standardised errors}
\label{app:proj}

Likelihood theory tells us that the asymptotic distribution of the maximum likelihood estimator $\hat{\latent}$ of the true parameters $\latent$ is
\begin{equation*}
  \hat{\latent} \sim \normal \left[ \latent, I(\latent)^{-1} \right]
\end{equation*}
where $I(\latent)$ is the expected information matrix.
The information matrix can be computed numerically as part of the optimisation procedure used to find $\hat{\theta}$.

Therefore, given $\hat{\latent}$ and equivalent \cotwo\ forcings $\forcing(t)$ for the historical and RCP4.5 experiments, we can sample the full projection uncertainty due to natural variability and parameter uncertainty for each fitted model as follows:
\begin{itemize}
  \item Sample $\latent^\star \sim \normal \left[ \hat{\latent}, I(\hat{\latent})^{-1} \right]$;
  \item Let $\state^\star(0) = \left( 0, 0, 0, 0 \right)^\prime$, i.e., pre-industrial equilibrium;
  \item For $t$ in $1,\ldots,T$
  \begin{itemize}
    \item Sample $\state^\star(t)$ from Equation~\ref{eqn:evolution_eqn}, conditional on $\state^\star(t-1)$, $\forcing(t)$ and $\latent^\star$;
    \item Sample $\vect{\out}^\star(t)$ from Equation~\ref{eqn:forecast_eqn}, conditional on $\state^\star(t)$ and $\latent^\star$;
  \end{itemize}
\end{itemize}
By repeating the sampling procedure we can obtain as many samples $\vect{\out}^\star(1),\ldots,\vect{\out}^\star(T)$ as we wish, sampling the full extent of both the parameter uncertainty and natural variability.

The standardised prediction errors at each time $t$ are then defined as
\begin{equation*}
  \frac{Y(t) - \E \left[ \out^\star(t) \right]}{\sqrt{\V \left[ \out^\star(t) \right]}}
\end{equation*}
where $Y(t)$ is the climate model output for the historical/RCP4.5 scenario at time $t$ and $\E \left[ \out^\star(t) \right]$ and $\V \left[ \out^\star(t) \right]$ are estimated from the samples $\out^\star(t)$.

\section{Extended state space representation}
\label{app:full}

The state-space representation of the extended model given by Equations~\ref{eqn:forcing}--\ref{eqn:disc} conditioned on the shared discrepancy $\shared(t)$ is given by
\begin{align*}
  \evolutionreg & = 
  \begin{bmatrix}
   \gamma F_C & \gamma F_V & 0 \\
   0 & 0 & 0 \\
   0 & 0 & 0 \\
   0 & 0 & 0 \\
   0 & 0 & 1
  \end{bmatrix} &
  \vect{\forcing}(t) = 
  \begin{bmatrix}
    \forcing_C(t) \\ \forcing_V(t) \\ \shared(t)
  \end{bmatrix}  
\end{align*}
and
\begin{align*}
  \evolution & =
  \begin{bmatrix}
    -\gamma & 0 & 0 & 0 & 0 \\
    \frac{1}{C_1} & -\frac{k_1 + k_2}{C_1} & \frac{k_2}{C_1} & 0 & \frac{1}{C_1} \\
    0 & \frac{k_2}{C_2} & -\frac{k_2 + \varepsilon k_3}{C_2} &  \frac{\varepsilon k_3}{C_2} & 0 \\
    0 & 0 & \frac{k_3}{C_3} & -\frac{k_3}{C_3} & 0 \\
    0 & 0 & 0 & 0 & 0 \\
  \end{bmatrix} &
  \evolutionvar & =
  \begin{bmatrix}
    \sigma_F^2 & 0 & 0 & 0 & 0 \\
    0 & \frac{\sigma_T^2}{C_1^2} & 0 & 0 & 0 \\
    0 & 0 & 0 & 0 & 0 \\
    0 & 0 & 0 & 0 & 0 \\
    0 & 0 & 0 & 0 & \sigma_\disc^2 \\
  \end{bmatrix}.
\end{align*}
The discretised version is
\begin{align*}
  \forecast_d &= 
  \begin{bmatrix}
    0 & 1 & 0 & 0 & 0 \\
    1 & -k_1 & (1 - \varepsilon) k_3 & -(1 - \varepsilon) k_3 & 1
  \end{bmatrix} &
  \forecastvar_d &= 
  \begin{bmatrix}
    0 & 0 \\
    0 & 0
  \end{bmatrix}
\end{align*}
and
\begin{align*}
  \evolution_d    & = 
    e^\evolution &
  \evolutionreg_d & = 
    \evolution^{-1} (\evolution_d - \mat{I}) \evolutionreg &
  \evolutionvar_d & = 
    \int_{s=0}^1 e^{\evolution s} \evolutionvar e^{\evolution^\prime s}.
\end{align*}

\bibliographystyle{agu}
\bibliography{library}

\end{document}


\renewcommand{\theequation}{S\arabic{equation}}
\renewcommand{\thesection}{S\arabic{section}}

\maketitle

\tableofcontents

\section{Prior specifications}
\label{sec:priors}

We need to specify priors for $\vect{\mu}_\latent$, $\mat{\Sigma}_\latent$ and $\sigma_\shared$.
The obvious conjugate choice of priors for $\vect{\mu}_\latent$ and $\mat{\Sigma}_\latent$ is
\begin{align}
  \vect{\mu}_\latent & \sim \normal \left( \vect{\mu}_0, \mat{\Sigma}_0 \right)   &
  \mat{\Sigma}_\latent & \sim \invwishart \left( \mat{\Psi}, d \right)
  \label{eqn:param-prior}
\end{align}
where $\vect{\mu}_0$ is a known vector of length 13, $\mat{\Sigma}_0$ and $\mat{\Psi}$ are known $13 \times 13$ positive-definite matrices, and $d \geq 13$ is a known real-valued scalar.
We specify an informative prior on the $\log$ scale for $\vect{\mu}_\latent$  based loosely on the individual model analysis of \citet{Cumminsetal2020} with mean vector
\begin{align*}
  \vect{\mu}_0 = \log \left( 2.0,5.0,20,100,1.0,2.0,1.0,0.5,0.5,3.0,20,0.05\right)^\prime
\end{align*}
and where the covariance $\mat{\Sigma}_0$ is diagonal with $[\log(10)/3]^2$ on the diagonal and zeros on all the off-diagonal elements.
This effectively constrains the order of magnitude of the mean over the models of each EBM parameter, e.g., $\log(\mu_\gamma) \sim \normal \left[ \log(2.0), (\log(10)/3)^2 \right]$ implies that $\mu_\gamma$ lies between 0.2 and 20 with probability 0.997, with equal probability of lying between 0.2 and 2.0, and 2.0 and 20, and probability 0.633 of lying between 1.0 and 4.0.
The prior range for the mean is always greater than the range over the models found by \citet{Cumminsetal2020}, so the prior is only moderately informative.
For $\mat{\Sigma}_\latent$, we specify a vague prior with degrees-of-freedom $d = 13$, and where the scale matrix $\mat{\Psi}$ is diagonal with $1\,000$ on the diagonal, zeros on all the off-diagonal elements.

The conjugate prior for $\sigma_\shared^2$ is 
$\sigma_\shared^2 \sim \invgammadist \left( a, b \right)$
where $a$ and $b$ are known positive real scalars.
However, in practice we prefer
\begin{align}
  \log \left( \sigma_\shared \right) \sim \normal \left( a, b \right)
  \label{eqn:sigma-prior}
\end{align}
where $a$ and $b > 0$ are known real scalars
This allows us to specify an informative prior for the order of magnitude of $\sigma_\shared$ where $a = \log(0.1)$ and $b = [\log(10)/3]^2$, i.e., $\sigma_\shared$ is constrained to lie between 0.01 and 1.0 with probability 0.997.


\section{Posterior inference}


In Section~\ref{sec:inference} of the main text, we assumed that the joint posterior had the separable form 
\begin{multline*}
 \Pr \left( \latent_\obs, \shared, \latent_1,\ldots,\latent_M, \vect{\mu}_\latent, \mat{\Sigma}_\latent, \sigma_\shared \mid \obs_H, \sigma_\obs^2, \kappa, \data \right) 
   = \\ 
   \Pr \left( \latent_\obs \mid \shared, \vect{\mu}_\latent, \mat{\Sigma}_\latent, \kappa, \obs_H, \sigma_\obs^2, X_H \right)
   \Pr \left( \shared, \latent_1,\ldots,\latent_M, \vect{\mu}_\latent, \mat{\Sigma}_\latent, \sigma_\shared \mid \data \right).
\end{multline*}
Therefore, the real world parameters $\latent_\obs$ can be inferred separately, conditional on the shared discrepancy $\shared$ and the common parameters $\vect{\mu}_\latent$ and $\mat{\Sigma}_\latent$.

\subsection{Sampling the parameters of the real world}

The posterior of the real world parameters $\latent_\obs$ is proportional to 
\begin{align}
  \Pr \left( \latent_\obs \mid \vect{\mu}_\latent, \mat{\Sigma}_\latent, \kappa, \shared, \obs_H, \sigma_Z^2, \vect{\forcing}_H \right)
    \propto \Pr \left( \obs_H \mid \latent_\obs, \shared, \sigma_Z^2, \vect{\forcing}_H \right)
            \Pr \left( \latent_\obs \mid \vect{\mu}_\latent, \mat{\Sigma}_\latent, \kappa \right)            
  \label{eqn:obs-post-prod}
\end{align}
where the prior $\Pr \left( \latent_\obs \mid \vect{\mu}_\latent, \mat{\Sigma}_\latent, \kappa \right)$ is given by Equation~\ref{eqn:obs_par}, and the likelihood $\Pr \left( \obs_H \mid \latent_\obs, \shared, \vect{\forcing}_H \right)$ can be evaluated using the state-space representation in Appendix~\ref{app:full} and the Kalman filter in Appendix~\ref{app:disc}.
We sample the posterior using a Metropolis-Hastings step with multivariate normal proposal
\begin{align}
  \log \left( \latent_\obs^\star \right) & \sim \normal \left[ \log \left( \latent_\obs \right), \mat{S}_\obs \right]
  \label{eqn:sample-obs}
\end{align} 
where the proposal covariance $\mat{S}_\obs$ is optimised during the burn-in period to give an acceptance probability of approximately $0.26$ using robust adaptive Metropolis-Hastings \citep{Vihola2012}.
The proposal covariance $\mat{S}_\obs$ is then fixed for the sampling period.
Since the proposal distribution in Equation~\ref{eqn:sample-obs} and the prior in Equation~\ref{eqn:obs_par} are both on the log scale, no Jacobian adjustment is necessary.

\subsection{Sampling the model specific and common parameters}

As noted in Section~\ref{sec:inference} of the main text, the inclusion of the shared discrepancy $\shared(t)$ means that the model specific parameters $\latent_1,\ldots,\latent_M$ have a complicated joint likelihood function that is difficult and expensive to evaluate.
If we condition on the shared discrepancy $\shared = \lbrace \shared(1),\ldots,\shared(\tau_F)\rbrace$ , then $\latent_1,\ldots,\latent_M$ are independent given the ensemble parameters $\vect{\mu}_\latent$ and $\mat{\Sigma}_\latent$ (Equation~\ref{eqn:params}).
However, sampling $\shared$ would still require us to be able to evaluate the joint likelihood function.
This can be avoided by noting that $\shared(t)$ only enters the model through the individual discrepancies $\disc(t)$ in Equation~\ref{eqn:disc}.
So, if we can sample the individual discrepancies $\disc_m = \lbrace \disc_m(1),\ldots,\disc_m(\tau_F) \rbrace$ ($m = 1,\ldots,M$), then we only require the joint likelihood of $\disc_1,\ldots,\disc_M$ and $\shared$ 
The full posterior is then
\begin{equation}
  \Pr \left( \latent_1,\ldots,\latent_M, \vect{\mu}_\latent, \mat{\Sigma}_\latent, \shared, \sigma_\shared \mid \data \right) 
  = \int \Pr \left( \latent_1,\ldots,\latent_M, \vect{\mu}_\latent, \mat{\Sigma}_\latent, \disc_1,\ldots,\disc_M, \shared, \sigma_\shared \mid \data \right) \dd \disc_1,\ldots,\disc_M
  \label{eqn:full-posterior}
\end{equation}
where
\begin{align*}
  \data = \left( \vect{\out}_{A1},\ldots,\vect{\out}_{AM}, \vect{\out}_{H1},\ldots,\vect{\out}_{HM}, \vect{\out}_{F1},\ldots,\vect{\out}_{FM}, \vect{\forcing}_A, \vect{\forcing}_H, \vect{\forcing}_F \right)
\end{align*}
is the available data, where $\vect{\out}_{Am}$ represents the bivariate time series $\vect{\out}_{Am}(t) = \left[ T_1(t), N(t) \right]^\prime$ ($t = 1,\ldots,\tau_A$) of output from climate model $m$ ($m = 1,\ldots,M$) for the \abrupt\ experiment, $\vect{\out}_{Hm}$ represents the same from the historical experiment ($t = 1,\ldots,\tau_H$), and $\vect{\out}_{Fm}$ from the RCP4.5 experiment ($t = \tau_H+1,\ldots,\tau_F$).
In addition, $\vect{\forcing}_A$ is the time series $\vect{\forcing}_A(t) = \left[ \forcing_C(t),\forcing_V(t) \right]^\prime$ of forcings for the \abrupt\ experiment, $\vect{\forcing}_H$ represents the same from the historical experiment ($t = 1,\ldots,\tau_H$), and $\vect{\forcing}_F$ from the RCP4.5 experiment ($t = \tau_H+1,\ldots,\tau_F$).
Note that volcanic forcing is set to zero ($\forcing_V(t) = 0$) during the \abrupt\ and RCP4.5 experiments.


The full posterior in Equation~\ref{eqn:full-posterior} can be sampled by Gibb's sampling, iterating over the full conditional distributions of the individual elements.
The simplified form of the full conditionals after removing unnecessary dependencies is
\begin{align}
  & \Pr \left( \latent_m \mid \vect{\mu}_\latent, \mat{\Sigma}_\latent, \shared, \vect{\out}_{Am}, \vect{\out}_{Hm}, \vect{\out}_{Fm}, \vect{\forcing}_{Am}, \vect{\forcing}_{Hm}, \vect{\forcing}_{Fm} \right) & m & = 1,\ldots,M
  \label{eqn:latent-post} \\
  & \Pr \left( \disc_m \mid \latent_m, \shared, \vect{\out}_{Hm}, \vect{\out}_{Fm}, \vect{\forcing}_{Hm}, \vect{\forcing}_{Fm} \right) & m & = 1,\ldots,M
  \label{eqn:disc-post} \\
  & \Pr \left( \shared \mid \latent_1,\ldots,\latent_M, \disc_1,\ldots,\disc_M, \sigma_\shared \right)
  \label{eqn:shared-post} \\  
  & \Pr \left( \vect{\mu}_\latent \mid \latent_1,\ldots,\latent_M, \mat{\Sigma}_\latent \right)
  \label{eqn:mu-post} \\
  & \Pr \left( \mat{\Sigma}_\latent \mid \latent_1,\ldots,\latent_M, \vect{\mu}_\latent \right)
  \label{eqn:Sigma-post} \\
  & \Pr \left( \sigma_\shared \mid \shared \right).
  \label{eqn:sigma-post}
\end{align}
Equations~\ref{eqn:latent-post}--\ref{eqn:sigma-post} strictly describe a partially collapsed Gibb's sampler \citep{VanDyk2008}, since we do not condition on $\disc_m$ in Equations~\ref{eqn:latent-post}.
Marginalising over $\disc_m$ enables more efficient sampling of $\latent_m$, specifically $\sigma_{\disc m}$.
The partially collapsed sampler will maintain the correct stationary distribution provided that the model specific parameters $\latent_m$ ($m=1,\ldots,M$) are sampled before the discrepancies $\delta_m$ ($m=1,\ldots,M$).

\subsubsection{Sampling the model specific parameters}

The posterior of the model specific parameters $\latent_1,\ldots,\latent_M$ in Equation~\ref{eqn:latent-post} is proportional to
\begin{multline}
  \Pr \left( \latent_m \mid  \vect{\mu}_\latent, \mat{\Sigma}_\latent, \shared, \vect{\out}_{Am}, \vect{\out}_{Hm}, \vect{\out}_{Fm}, \vect{\forcing}_A, \vect{\forcing}_H, \vect{\forcing}_F \right) \\
  \propto \Pr \left( \latent_m \mid \vect{\mu}_\latent, \mat{\Sigma}_\latent \right)
  \Pr \left( \vect{\out}_{Am} \mid \latent_m, \vect{\forcing}_A \right)
  \Pr \left( \vect{\out}_{Hm}, \vect{\out}_{Fm} \mid \latent_m, \shared,  \vect{\forcing}_H, \vect{\forcing}_F \right)
  \label{eqn:latent-post-prod}
\end{multline}
where the prior $\Pr \left( \latent_m \mid \vect{\mu}_\latent, \mat{\Sigma}_\latent \right)$ can be evaluated from Equation~\ref{eqn:params}, and the likelihoods $\Pr \left( \vect{\out}_{Am} \mid \latent_m, \vect{\forcing}_A \right)$ of the \abrupt\ and $ \Pr \left( \vect{\out}_{Hm}, \vect{\out}_{Fm}  \mid \latent_m, \shared, \vect{\forcing}_H, \vect{\forcing}_F \right)$ of the historical and future model outputs can be evaluated via the Kalman filter using Equation~\ref{eqn:likelihood}.
We sample the posterior using Metropolis-Hastings steps with model specific multivariate normal proposals
\begin{align}
  \log \left( \latent_m^\star \right) & \sim \normal \left[ \log \left( \latent_m \right), \mat{S}_m \right] & m & = 1,\ldots,M
  \label{eqn:sample-latent}
\end{align} 
where $\latent_m^\star$ is the proposed value and the proposal variances $\mat{S}_m$ are optimised during the burn-in period to give acceptance probabilities of approximately $0.26$ using robust adaptive Metropolis-Hastings \citep{Vihola2012}.
The proposal covariances $\mat{S}_m$ are then fixed for the sampling period.
Since the proposal distribution in Equation~\ref{eqn:sample-latent} and the prior in Equation~\ref{eqn:params} are both on the log scale, no Jacobian adjustment is necessary.

\subsubsection{Sampling the model specific discrepancies}
\label{sec:sampling-disc}

The model specific forcing discrepancies $\disc_m$ ($m=1,\ldots,M$) are part of the state vector $\state(t)$ of each EBM representation.
Conditional on the shared discrepancy $\shared$ and the model specific parameters $\latent_m$, each EBM representation is independent of all the others, so we can sample the discrepancies $\disc_m$ one at a time.
In order to sample the individual discrepancy $\disc_m = \lbrace \disc_m(1),\ldots,\disc_m(\tau_F) \rbrace$ of model $m$ we need to be able to sample $\Pr \left[ \state(1),\ldots,\state(\tau_F) \mid \latent_m, \shared, \vect{\out}_{Hm},\vect{\out}_{Fm},\vect{\forcing}_H,\vect{\forcing}_F \right]$, i.e., the joint distribution of the state given all the output from that model.
This can be achieved by the forward filtering, backward sampling algorithm \citep{Fruhwirth1994}.
Forward filtering refers to Kalman filtering (Equations~\ref{eqn:state_pred}--\ref{eqn:state_post}).
Backward sampling relies on the following decomposition of the joint distribution of the state
\begin{equation}
  \begin{split}
    \Pr \left[ \state(1),\ldots,\state(T) \mid \data(1:T) \right]
     & = \Pr \left[ \state(T) \mid \data(1:T) \right]
         \Pr \left[ \state(T-1) \mid \state(T),\data(1:T-1) \right]
         \ldots \\
     & \quad \ldots 
       \Pr \left[ \state(1) \mid \state(2),\data(1) \right]     
       \Pr \left[ \state(0) \mid \state(1),\data(0) \right].
  \end{split}
  \label{eqn:back-decomp}
\end{equation}
where $\data(1:t) = \data(1),\ldots,\data(t)$ represents all the available data up to time $t$ ($t = 1,\ldots,T$) and $\data(0)$ is the prior for the state at time $t=0$.
It can be shown that 
\begin{equation}
  \left[ \state(t) \mid \state(t+1),\data(1:t) \right]
    \sim \normal \left[ \fullpostexpstate(t), \fullpostvarstate(t) \right]
  \label{eqn:back-dist}
\end{equation}
\citep[Chapter~15]{WestHarrison} where
\begin{align}
  \fullpostexpstate(t) & = \postexpstate(t) + 
    \backkalmangain(t) \left[ \state(t+1) - \priorexpstate(t+1) \right] & 
  \fullpostvarstate(t) & = \postvarstate(t) -
    \backkalmangain(t) \priorvarstate(t+1) \backkalmangain^\prime(t)
  \label{eqn:back-sample}
\end{align}
and $\backkalmangain(t) = \postvarstate(t) \evolution_d^\prime(t+1) \priorvarstate^{-1}(t+1)$ is the backward gain, and the quantities $\priorexpstate(t)$, $\priorvarstate(t)$, $\postexpstate(t)$ and $\postvarstate(t)$ are as defined in the Kalman filter (Equations~\ref{eqn:state_pred}--\ref{eqn:state_post}).

So in order to sample the discrepancy $\disc$ of model $m$
\begin{itemize}
  \item Use the Kalman filter to obtain the sequences of prior distributions $\left[ \state_m(t) \mid \data_m(1:t-1), \latent_m \right] \sim \normal \left[ \priorexpstate(t), \priorvarstate(t) \right]$ and posterior distributions $\left[ \state_m(t) \mid \data_m(1:t), \latent_m \right] \sim \normal \left[ \postexpstate(t), \postvarstate(t) \right]$ where $\data_m(1:t) = \lbrace \vect{\out}_{Hm}(1:t), \vect{\out}_{Fm}(1:t), \vect{\forcing}_H(1:t), \vect{\forcing}_F(1:t), \shared(1:t) \rbrace$ and $t = 1,\ldots,\tau_F$;
  \item Sample $\left[ \state_m^\star(\tau_F) \mid \data_m(1:\tau_F), \latent_m \right] \sim \normal \left[ \postexpstate(\tau_F), \postvarstate(\tau_F) \right]$;
  \item For $t = \tau_F-1,\ldots,1$:
  \begin{itemize}
    \item Sample $\left[ \state_m^\star(t) \mid \state_m^\star(t+1),\data_m(1:t) \right]
    \sim \normal \left[ \fullpostexpstate(t), \fullpostvarstate(t) \right]$ using Equation~\ref{eqn:back-sample}.
  \end{itemize}
\end{itemize}
Once we have the sequence of samples $\state_m^\star(1),\ldots,\state_m^\star(\tau_F)$, simply extract the discrepancy sequence $\disc_m^\star = \disc_m^\star(1),\ldots,\disc_m^\star(\tau_F)$, and discard the rest of the state.

\subsubsection{Sampling the shared discrepancy}
\label{sec:sampling-shared}

In discrete time, Equations~\ref{eqn:disc} and \ref{eqn:shared}, describe a $\VAR(1)$ model for the vector of model specific discrepancies $\vect{\disc}(t)  = \left[ \disc_1(t),\ldots,\disc_M(t) \right]^\prime$ such that
\begin{align}
\vect{\disc}(t) & = \evolution_\disc \vect{\disc}(t-1) + \vect{w}_\disc(t) &
\vect{w}_\disc(t) & \sim \normal \left( \vect{0} , \mat{\evolutionvar}_\disc \right)
\end{align}
where
\begin{align}
\evolution_\disc & =
\begin{bmatrix}
  1      & 0      & \cdots & 0      \\
  0      & \ddots & \ddots & \vdots \\
  \vdots & \ddots & \ddots & 0      \\
  0      & \cdots & 0      & 1      
\end{bmatrix} &
\evolutionvar_\disc & = 
\begin{bmatrix}
  \sigma_\shared^2 + \sigma_{\delta 1}^2 & \sigma_\shared^2 & \cdots & \sigma_\shared^2 \\
  \sigma_\shared^2 & \ddots & \ddots & \vdots \\
  \vdots & \ddots & \ddots & \sigma_\shared^2 \\
  \sigma_\shared^2 & \cdots & \sigma_\shared^2 & \sigma_\shared^2 + \sigma_{\delta M}^2
\end{bmatrix}
\end{align}
are $M \times M$ matrices.
In order to sample the shared discrepancy $\shared(t)$ we form an augmented model
\begin{align}
 \vect{\disc}(t) & = \forecast_\disc \vect{\disc}^+(t)
 \label{eqn:disc_obs}
\end{align}
where $\vect{\disc}^+(t)  = \left[ \disc_1(t),\ldots,\disc_M(t),\shared(t) \right]^\prime$ and 
\begin{align}
\vect{\disc}^+(t) & = \evolution_\disc^+ \vect{\disc}^+(t-1) + \vect{w}_\disc^+(t) &
\vect{w}_\disc^+(t) & \sim \normal \left( \vect{0} , \mat{\evolutionvar}_\disc^+ \right)
  \label{eqn:disc_state}
\end{align}
with
\begin{align}
\forecast_\disc & =
\begin{bmatrix}
  \mat{I} & \vect{0}
\end{bmatrix} &
\evolution_\disc^+ & =
\begin{bmatrix}
  \evolution_\disc & \vect{0} \\
  \vect{0}          & 0
\end{bmatrix} &
\evolutionvar_\disc^+ & = 
\begin{bmatrix}
  \evolutionvar_\disc & \vect{\sigma}_\shared^2 \\
  \vect{\sigma}_\shared^2 & \sigma_\shared^2
\end{bmatrix}
  \label{eqn:disc_mat}
\end{align}
and $\mat{I}$ is the $M \times M$ identity matrix.
Equations~\ref{eqn:disc_obs}, \ref{eqn:disc_state}, \ref{eqn:disc_mat} describe a state-space model (Equations~\ref{eqn:forecast_eqn} and \ref{eqn:evolution_eqn}) with zero observational error $\forecastvar$ and no predictors $\vect{\forcing}(t)$.
Therefore, we can sample the shared discrepancy $\shared$ by forward filtering, backward sampling as follows:
\begin{itemize}
  \item Use the Kalman filter to obtain the sequences of prior distributions $\left[ \vect{\disc}^+(t) \mid \data(1:t-1), \sigma_{\disc 1},\ldots,\sigma_{\disc M}, \sigma_\shared \right] \sim \normal \left[ \priorexpstate(t), \priorvarstate(t) \right]$ and posterior distributions $\left[ \vect{\disc}^+(t) \mid \data(1:t), \sigma_{\disc 1},\ldots,\sigma_{\disc M}, \sigma_\shared \right] \sim \normal \left[ \postexpstate(t), \postvarstate(t) \right]$ where $\data(1:t) = \lbrace \disc_1(1:t),\ldots,\disc_M(1:t), \shared(1:t) \rbrace$ and $t = 1,\ldots,\tau_F$;
  \item Sample $\left[ \vect{\disc}^{+\star}(\tau_F) \mid \data(1:\tau_F), \sigma_{\disc 1},\ldots,\sigma_{\disc M}, \sigma_\shared \right] \sim \normal \left[ \postexpstate(\tau_F), \postvarstate(\tau_F) \right]$;
  \item For $t = \tau_F-1,\ldots,1$:
  \begin{itemize}
    \item Sample $\left[ \vect{\disc}^{+\star}(t) \mid \vect{\disc}^{+\star}(t+1),\data(1:t) \right]
    \sim \normal \left[ \fullpostexpstate(t), \fullpostvarstate(t) \right]$ using Equation~\ref{eqn:back-sample}.
  \end{itemize}
\end{itemize}
Once we have the sequence of samples $\vect{\disc}^{+\star}(1),\ldots,\vect{\disc}^{+\star}(\tau_F)$, simply extract the discrepancy sequence $\shared^\star = \shared^\star(1),\ldots,\shared^\star(\tau_F)$, and discard the rest of the sampled state.
Since there is no observational error $\forecastvar$, we have $\disc_m^\star(t) = \disc_m(t)$ for all $t = 1,\ldots,\tau_F$ and $m = 1,\ldots,M$, i.e., the values of the individual discrepancies sampled from the augmented model must equal the values sampling from the full sampled from the extended state space model in the previous section.

\subsubsection{Sampling the distribution over the models}
\label{sec:sampling-params}

The hierarchical parameters $\vect{\mu}_\latent$ and $\mat{\Sigma}_\latent$ have conjugate prior distributions in Equation~\ref{eqn:param-prior} so can be sampled directly from the full conditionals in Equations~\ref{eqn:mu-post} and \ref{eqn:Sigma-post}
\begin{align*}
  \Pr \left( \vect{\mu}_\latent, \mid \mat{\Sigma}_\latent,  \latent_1,\ldots,\latent_M \right)
    & \propto \Pr \left( \vect{\mu}_\latent \right) 
              \prod_{m=1}^M \Pr \left( \latent_m \mid \vect{\mu}_\latent, \mat{\Sigma}_\latent \right) \\
  \Pr \left( \mat{\Sigma}_\latent, \mid \vect{\mu}_\latent,  \latent_1,\ldots,\latent_M \right)
    & \propto \Pr \left( \mat{\Sigma}_\latent \right) 
              \prod_{m=1}^M \Pr \left( \latent_m \mid \vect{\mu}_\latent, \mat{\Sigma}_\latent \right)
\end{align*}
where $\Pr \left( \vect{\mu}_\latent \right)$ and $\Pr \left( \mat{\Sigma}_\latent \right)$ are the priors given by Equation~\ref{eqn:param-prior}, and $\Pr \left( \latent_m \mid \vect{\mu}_\latent, \mat{\Sigma}_\latent \right)$ is given by Equation~\ref{eqn:params}.
The sampling distributions are then
\begin{align*}
  \vect{\mu}_\latent^\star & \sim \normal \left[ \left( \mat{B}^{-1} + M \mat{\Sigma}_\latent^{-1} \right)^{-1} \left( \mat{B}^{-1} \vect{a} + \mat{\Sigma}_\latent^{-1} \sum_{m = 1}^M \latent_m \right), \left( \mat{B}^{-1} + M \mat{\Sigma}_\latent^{-1} \right)^{-1} \right] \\
  \mat{\Sigma}_\latent^\star & \sim \invwishart \left[ \mat{C} + \sum_{m=1}^M \left( \latent_m - \vect{\mu}_\latent \right) \left( \latent_m - \vect{\mu}_\latent \right)^\prime, d + M \right].
\end{align*}

\subsubsection{Sampling the variance of the shared discrepancy}

The posterior in Equation~\ref{eqn:sigma-post} is proportional to
\begin{align}
  \Pr \left( \sigma_\shared \mid \shared \right) & \propto \Pr \left( \shared \mid \sigma_\shared \right) \Pr \left( \sigma_\shared \right)
  \label{eqn:sigma-post-prod}
\end{align}
where the prior $\Pr \left( \sigma_\shared \right)$ is given by Equation~\ref{eqn:sigma-prior} and the likelihood $\Pr \left( \shared \mid \sigma_\shared \right)$ is given by Equation~\ref{eqn:shared}.
We sample the posterior distribution using a Metropolis-Hastings step with a normal proposal
\begin{align}
  \log \left( \sigma_\shared^\star \right) & \sim \normal \left[ \log \left( \sigma_\shared \right), s_\shared^2 \right]
  \label{eqn:sample-sigma}
\end{align} 
where the proposal variance $s_\shared^2$ is optimised during the burn-in period to give an acceptance probability of approximately $0.44$ using robust adaptive Metropolis-Hastings \citep{Vihola2012}.
The proposal variance $s_\shared^2$ is then fixed for the sampling period.
Since the proposal distribution in Equation~\ref{eqn:sample-sigma} and the prior in Equation~\ref{eqn:sigma-prior} are both on the log scale, no Jacobian adjustment is necessary.

\subsection{Initialisation}
\label{sec:initialisation}

The model specific EBM parameters $\latent_m$ ($m=1,\ldots,M$) are intialised by sampling from
\begin{align*}
  \latent_m^\star & \sim \normal \left( \vect{\mu}_m, \mat{\Sigma}_m \right) & m &= 1,\ldots,M
\end{align*}
where $\vect{\mu}_m$ is the maximum a posteriori (MAP) estimate found by numerical maximisation of the posterior in Equation~\ref{eqn:latent-post-prod}, and $\mat{\Sigma}_m = I(\latent_m)^{-1}$ is the inverse of the expected information matrix evaluated at the MAP estimate.
The initial posteriors for all $\latent_m$ ($m=1,\ldots,M$) are conditioned on $\shared(t) = 0$ for all $t = 1,\ldots,\tau_F$, i.e., no shared discrepancy, and the prior $\Pr \left( \latent_m \mid \vect{\mu}_\latent, \mat{\Sigma}_\latent \right)$ is evaluated at $\vect{\mu}_0$ and $\mat{\Sigma_0}$ from Section~\ref{sec:priors}.
The initial estimates of the proposal distributions $\mat{S}_m$ ($m = 1,\ldots,M$) in Equation~\ref{eqn:sample-latent} are set to $\mat{S}_m = 2.38^2 \mat{\Sigma}_m / 13$, the optimal estimate of \citet{Roberts2009}.

The model specific discrepancies $\disc_m$ ($m = 1,\ldots,M$) are initialised by sampling as described in Section~\ref{sec:sampling-disc}, conditional on the initial $\latent_m^\star$ ($m=1,\ldots,M$) and $\shared(t) = 0$ for all $t = 1,\ldots,\tau_F$, i.e., no shared discrepancy.

The shared discrepancy $\shared$ is initialised by sampling as described in Section~\ref{sec:sampling-shared}, conditional on the initial $\disc_m^\star$ ($m = 1,\ldots,M$) and $\sigma_{\disc m}^\star$ ($m = 1,\ldots,M$) sampled as part of $\latent_m^\star$, and $\sigma_\shared$ is estimated by the standard deviation of the first differences of the mean over the $\disc_m^\star$, i.e., a random walk model for the mean.

The ensemble parameters $\vect{\mu}_\latent$ and $\mat{\Sigma}_\latent$ have conjugate priors and so are initialised by sampling as described in Section~\ref{sec:sampling-params}, conditional on the initial values of the model parameters $\latent_m^\star$ ($m=1,\ldots,M$).

The standard deviation of the shared discrepancy $\sigma_\shared$ is formally initialised by sampling from 
\begin{align*}
  \sigma_\shared^\star & \sim \normal \left( \mu_\sigma, \sigma_\sigma^2 \right)
\end{align*}
where $\mu_\sigma$ is the MAP estimate found by numerical maximisation of the posterior in Equation~\ref{eqn:sigma-post-prod} conditional on the initial estimate $\shared^\star$ of the shared discrepancy and the prior in Section~\ref{sec:initialisation}. 
The variance $\sigma_\sigma^2 = I(\sigma_\shared)^{-1}$ is the inverse of the expected information evaluated at the MAP estimate.
The initial estimate of the proposal variance $s_\shared^2$  in Equation~\ref{eqn:sample-sigma} is set to $s_\shared^2 = 2.38^2 \sigma_\sigma^2$.

Finally, the EBM parameters $\latent_\obs$ of the real world are intialised by sampling from
\begin{align*}
  \latent_\obs^\star & \sim \normal \left( \vect{\mu}_\obs, \mat{\Sigma}_\obs \right) & m &= 1,\ldots,M
\end{align*}
where $\vect{\mu}_\obs$ is the maximum a posteriori (MAP) estimate found by numerical maximisation of the posterior in Equation~\ref{eqn:obs-post-prod}, conditional on the initial values $\shared^\star$, $\vect{\mu}_\latent^\star$ and $\mat{\Sigma}_\latent^\star$ of the shared discrepancy and ensemble parameters.
The variance $\mat{\Sigma}_Z = I(\latent_Z)^{-1}$ is the inverse of the expected information matrix evaluated at the MAP estimate.
The initial estimate of the proposal distributions $\mat{S}_Z$ in Equation~\ref{eqn:sample-obs} is set to $\mat{S}_Z = 2.38^2 \mat{\Sigma}_Z / 13$.

\section{Sensitivity to choice of climate models}

In the analysis presented in the main text we included a subset of 13 of the available CMIP5 models in order to avoid biasing our inferences by including several versions of closely related models.
To test the dependence of our inferences on the choice of models, we ran the same analysis including all 24 models with the requried data.
The new models and their posterior parameter estimates are listed in Table~\ref{tab:parameters-all}.

\begin{figure}[t]
  \includegraphics[scale=1]{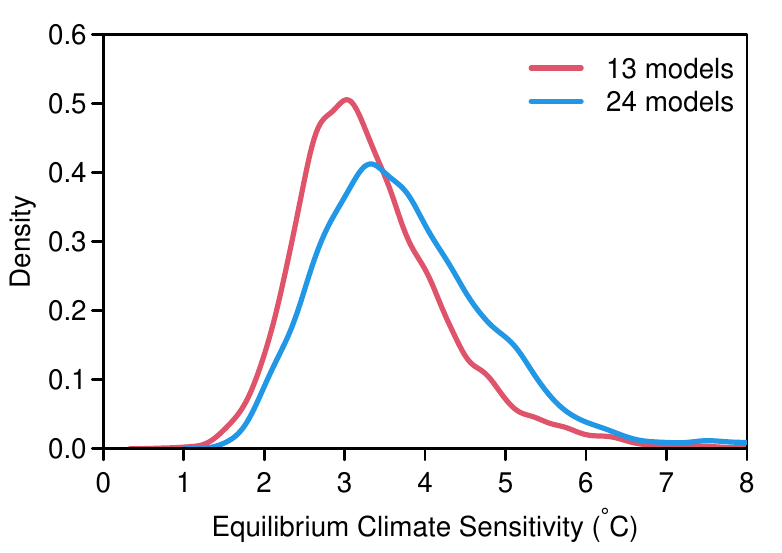}
  \caption{Equilibrium climate sensitivity. The posterior density of the equilibrium climate sensitivity in the real world under the RCP4.5 scenario with based on inference from 13 climate models (red) and 25 climate models (blue).}
  \label{fig:ecs-all}
\end{figure}

Figure~\ref{fig:ecs-all} compares the posterior distribution of ECS inferred using the 13 climate models described in the main text with the distribution inferred using all 25 models with the required outputs.
Using the 13 model subset, we estimate a median ECS of \SI{3.2}{\kelvin} with 90\,\% credible interval \SIrange{2.1}{5.1}{\kelvin}.
Using all 24 models, we estimate a median ECS of \SI{3.6}{\kelvin} with 90\,\% credible interval \SIrange{2.3}{5.8}{\kelvin}.
Equilibrium climate sensitivity is estimated as $F_C/k_1$.
In Table~\ref{tab:parameters-all}, the full ensemble contains additional variants of the two models in the original ensemble with the highest values of the \cotwo\ forcing coefficient $F_C$, and several other models with high values of $F_C$.
It also includes several variants and other models with very low values of $k_1$.
The overall effect is to shift the prior distribution for the real world towards higher values of ECS, as seen in Figure~\ref{fig:ecs-all}.
This highlights the importance of accounting for model dependence and the potential for biased inferences if it is ignored.

\begin{figure}[t]
  \includegraphics[scale=1]{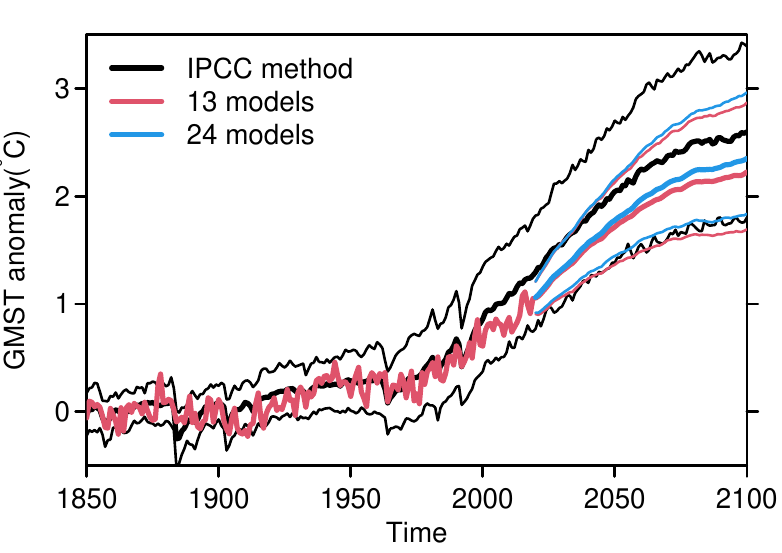}
  \caption{Global mean surface temperature anomaly projections. 
  Posterior predictive distribution for global mean surface temperature under  the RCP4.5 forcing scenario. 
  Thick lines indicate the posterior predictive means of the CMIP5 ensemble (black) and the real world based on 13 climate models (red) and the real world based on 25 climate models (blue). 
  Thin lines indicate marginal 90\,\% credible intervals.}
  \label{fig:projections-all}
\end{figure}

Figure~\ref{fig:projections-all} compares the projections of real world climate change under the RCP4.5 scenario inferred using the 13 climate models described in the main text, and all 24 models with the required outputs.
Using the 13 model subset, we estimate a median warming of \SI{1.5}{\kelvin} in 2081--2100 compared to 1986--2005, with 90\,\% credible interval \SIrange{1.1}{2.1}{\kelvin}.
Using all 24 available models, we estimate a median warming of \SI{1.6}{\kelvin} in 2081--2100 compared to 1986--2005, with 90\,\% credible interval \SIrange{1.2}{2.2}{\kelvin}.
The difference is not large, but is clearly visible in Figure~\ref{fig:projections-all}.

\begin{figure}[t]
  \includegraphics[scale=1]{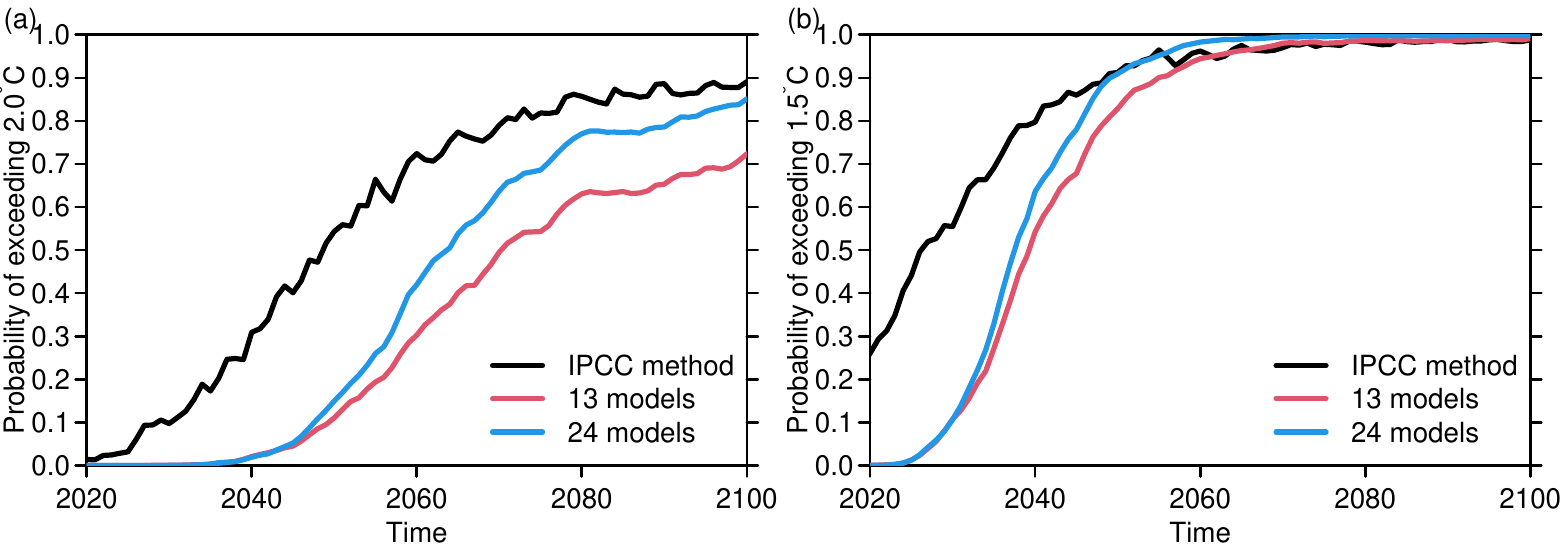}
  \caption{Probability of meeting the Paris agreement. 
 (a) The probability of climate change below $2.0\,^\circ\mathrm{C}$ under the RCP4.5 forcing scenario, and
 (b) the probability of climate change below $1.5\,^\circ\mathrm{C}$ under the RCP4.5 forcing scenario.}
  \label{fig:probabilities-all}
\end{figure}

The effects of the repeated models are even clearer in Figure~\ref{fig:probabilities-all} which compares the probabilities of meeting the targets set out in the Paris agreement under the RCP4.5 scenario inferred using the 13 climate models described in the main text, and all 24 models with the required outputs.
The probability of exdeeding \SI{2.0}{\kelvin} above the preindustrial average jumps from 0.72 using the 13 selected climate models to 0.85 when all 24 models are included.
This further emphasises the importance of accounting for model dependence in order to avoid biased estimates.

\section{Sensitivity to priors}

In Section~\ref{sec:priors} we specified moderately informative priors for the parameters $\vect{\mu}_\latent$ and $\sigma_\shared$.
In order to test the sensitivity of our inferences to those priors, we repeated the analysis with the following vague priors
\begin{align*}
  \vect{\mu}_\latent & \sim \normal \left( \vect{0}, \mat{\Sigma}_0 \right)  
\end{align*}
where the covariance $\mat{\Sigma}_0$ is diagonal with $1\,000$ on the diagonal and zeros on all the off-diagonal elements, and
\begin{align*}
  \log \left( \sigma_\shared \right) \sim \normal \left( 0, 1\,000 \right).
\end{align*}
Once again, a shorter MCMC run was used consisting of 4 chains each with a burnin of 25\,000 samples before 100\,000 final samples.

\begin{figure}[t]
  \includegraphics[scale=1]{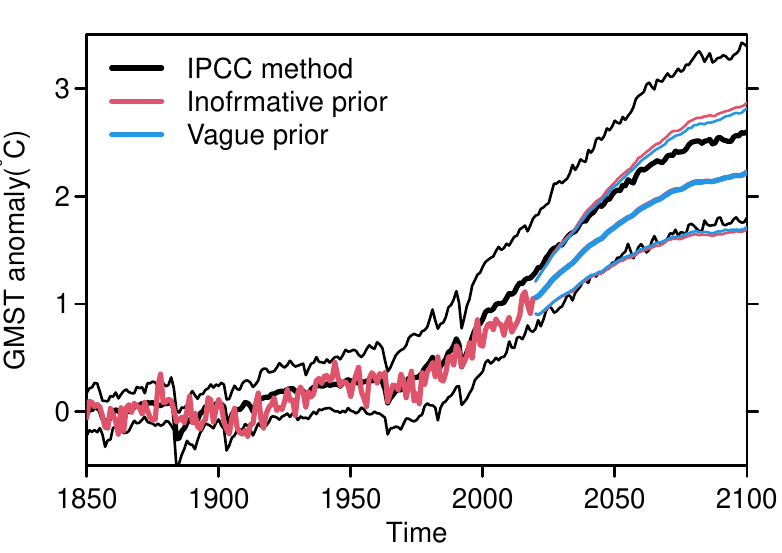}
  \caption{Global mean surface temperature anomaly projections. 
  Posterior predictive distribution for global mean surface temperature under  the RCP4.5 forcing scenario. 
  Thick lines indicate the posterior predictive means of the CMIP5 ensemble (black) and the real world with informative priors (red) and vague priors (blue). 
  Thin lines indicate marginal 90\,\% credible intervals.}
  \label{fig:projections-flat}
\end{figure}

Figure~\ref{fig:projections-flat} compares the projections of real world climate change under the RCP4.5 scenario inferred using moderately informative priors described in Section~\ref{sec:priors}, with vague uninformative priors.
Using the informative priors, we estimate a median warming of \SI{1.5}{\kelvin} in 2081--2100 compared to 1986--2005, with 90\,\% credible interval \SIrange{1.1}{2.1}{\kelvin}.
Using all the vague priors, we estimate a median warming of \SI{1.5}{\kelvin} in 2081--2100 compared to 1986--2005, with 90\,\% credible interval \SIrange{1.1}{2.1}{\kelvin}.
The projected future climate is entirely unchanged by the choice of priors.

\section{Sensitivity to coexchangeable coefficient}

\begin{figure}[t]
  \includegraphics[scale=1]{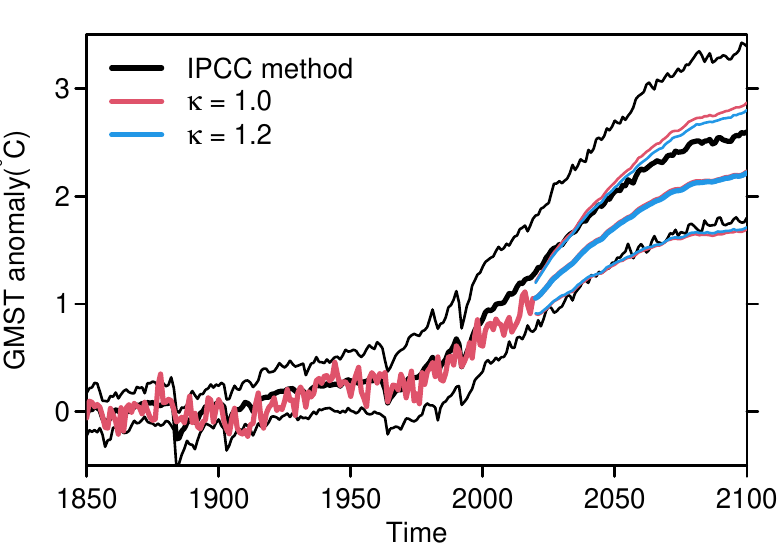}
  \caption{Global mean surface temperature anomaly projections. 
  Posterior predictive distribution for global mean surface temperature under  the RCP4.5 forcing scenario. 
  Thick lines indicate the posterior predictive means of the CMIP5 ensemble (black) and the real world with $\kappa = 1.0$ (red) and $kappa = 1.2$ (blue). 
  Thin lines indicate marginal 90\,\% credible intervals.}
  \label{fig:projections-kappa}
\end{figure}

Figure~\ref{fig:projections-kappa} compares the projections of real world climate change under the RCP4.5 scenario inferred under the standard exchangeable assumption that $\kappa = 1.0$, and an assumption of additional uncertainty in the real world with $\kappa = 1.2$.
Once again, a shorter MCMC run was used consisting of 4 chains each with a burnin of 25\,000 samples before 100\,000 final samples.
Using $\kappa = 1.0$, we estimate a median warming of \SI{1.5}{\kelvin} in 2081--2100 compared to 1986--2005, with 90\,\% credible interval \SIrange{1.1}{2.1}{\kelvin}.
Using $\kappa = 1.2$, we estimate a median warming of \SI{1.5}{\kelvin} in 2081--2100 compared to 1986--2005, with 90\,\% credible interval \SIrange{1.1}{2.1}{\kelvin}.
The projected future climate is unchanged by increasing the coexchangeable coefficient.

\section{The RCP8.5 scenario}
\label{sec:rcp85}

In addition to the RCP4.5 scenario in the main text, we also ran the same analysis for the stronger RCP8.5 scenario.
Table~\ref{tab:parameters-rcp85} shows the posterior parameter estimates based on the \abrupt\, historical and RCP8.5 scenarios for comparison with Table~\ref{tab:parameters} in the main text based on the \abrupt\, historical and RCP4.5 scenarios.
The parameter estimates are almost identical, emphasising that most of the information comes from the \abrupt\ experiments.
The future scenario might be expected to most strongly influence the individual discrepancy variances $\sigma_\delta$, but even these are very similar to the RCP4.5 estimates.

\begin{figure}[t]
  \includegraphics[scale=1]{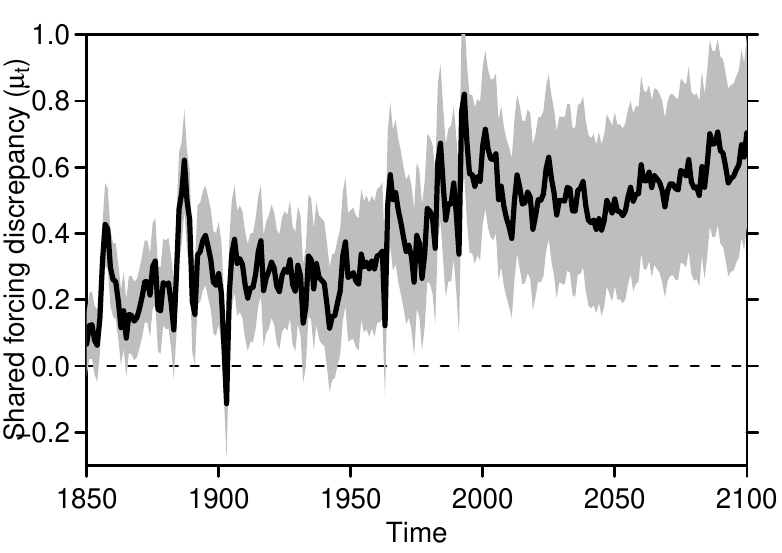}
  \caption{The shared forcing discrepancy. The posterior mean (solid line) and marginal 90\,\% credible intervals (shading) for the shared forcing discrepancy $\mu_t$ under the RCP8.5 forcing scenario.}
  \label{fig:shared-rcp85}
\end{figure}

The shared discrepancy $\mu(t)$ is plotted in Figure~\ref{fig:shared-rcp85} for comparison with Figure~\ref{fig:shared} of the main text.
As expected, the trajectories are essentially identical up to 2005, where the forcing changes from the historical to the future scenario.
The RCP4.5 and RCP8.5 scenarios diverge slowly at first, so the trajectories are similar up to 2020 when the observed period ends.
Where the RCP4.5 shared discrepancy declines slowly after 2020, the RCP8.5 shared discrepancy plateaus then slowly rises again.
This is likely to be a result of the stronger forcing inducing stronger feedbacks that are not accounted for by the simple EBM representations.

\subsection{Cross-validation}

\begin{figure}[t]
  \includegraphics[scale=1]{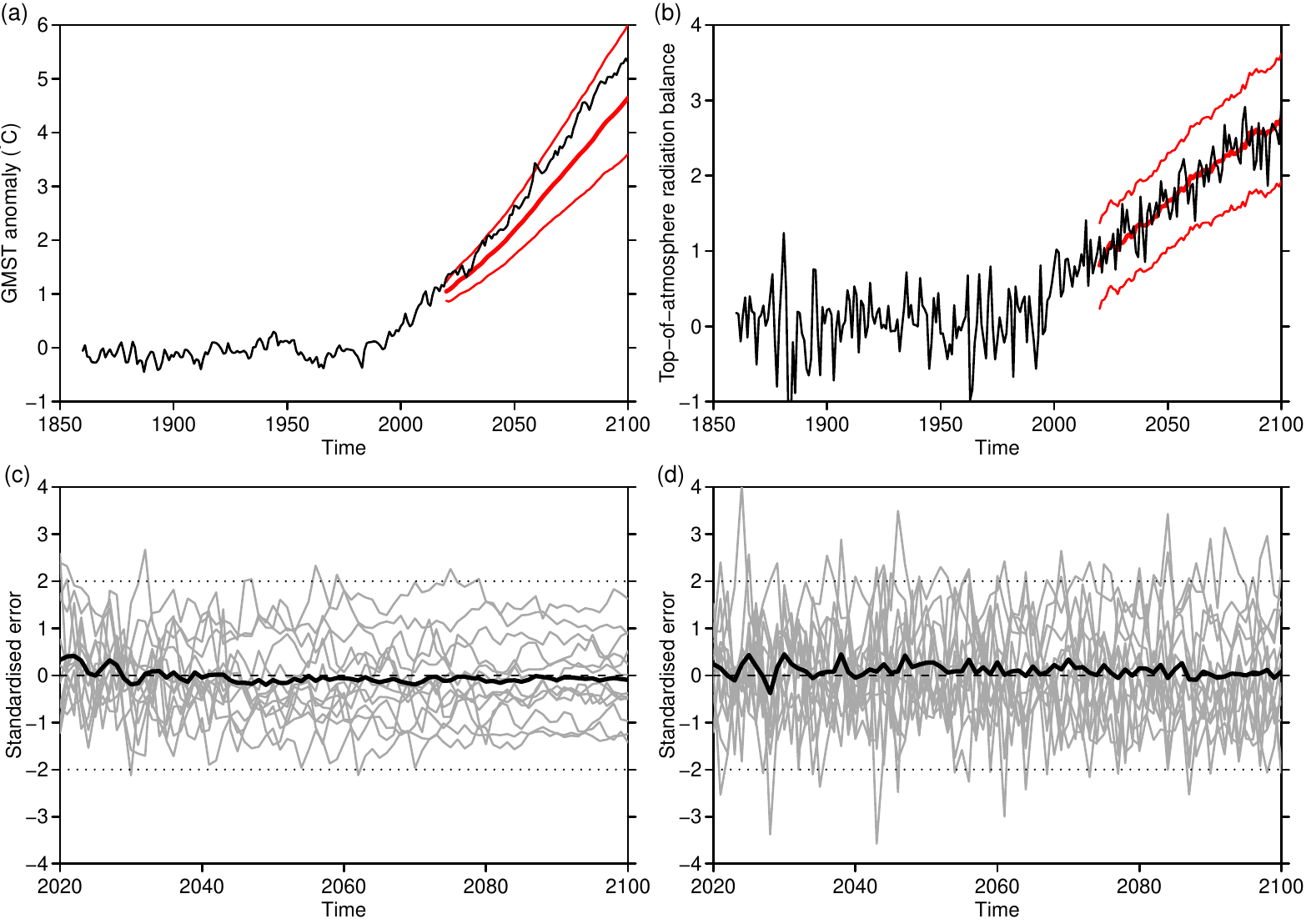}
  \caption{Cross validation.
           (a) and (b) global mean surface temperature and top-of-atmosphere radiation balance respectively in the HadGEM2-ES model under the RCP8.5 scenario.
           Black lines are the model simulations, thick red lines are the posterior predictive means based on the model including both individual and shared forcing discrepancies, and thin red lines are a marginal 90\,\% credible interval.
           (c) and (d) Standardised prediction errors for global mean surface temperature and top-of-atmosphere radiation balance respectively under the RCP~8.5 scenario including both individual and shared forcing discrepancies.
           Thin grey lines represent standardised errors from individual CMIP5 models.
           The thick black line is the ensemble mean.}
  \label{fig:cv-rcp85}
\end{figure}

The results of the cross-validation exercise under the RCP8.5 scenario are shown in Figure~\ref{fig:cv-rcp85} for comparison with Figure~\ref{fig:cv} of the main text.
Figure~\ref{fig:cv-rcp85}(a) and (b) once again show the example of the HadGEM2-ES model.
The results are qualitatively similar to the RCP4.5 scenario.
The credible intervals look plausible, almost entirely containing the excluded model output from 2020--2100.
The standardised errors in Figures~\ref{fig:cv-rcp85}(c) and (d) also tell a similar story to the RCP4.5 scenario.
There is no evidence of any residual bias.
Once again, reliability looks excellent, although the temperature projections may be slightly under-confident (credible intervals too wide).

\subsection{Equilibrium Climate Sensitivity}

\begin{figure}[t]
  \includegraphics[scale=1]{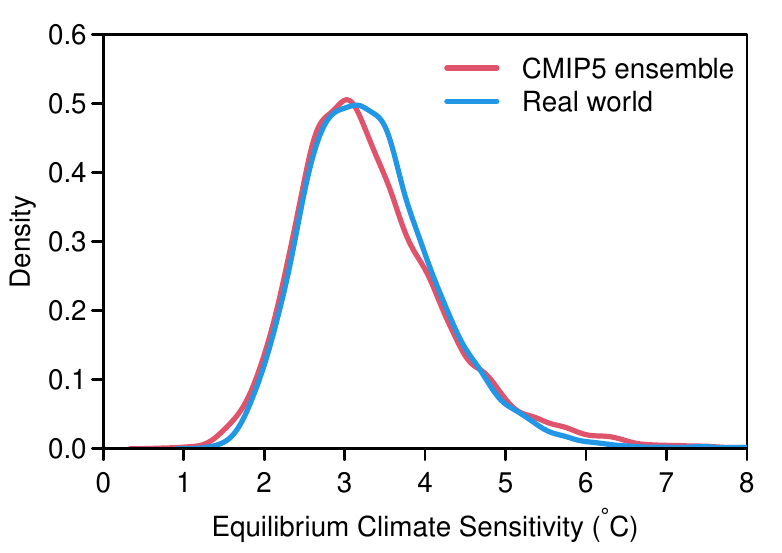}
  \caption{Equilibrium climate sensitivity. The posterior density of the equilibrium climate sensitivity for the real world (red) and the CMIP5 ensemble (black) under the RCP8.5 forcing scenario.}
  \label{fig:ecs-rcp85}
\end{figure}

Given the limited changes visible in Table~\ref{tab:parameters-rcp85} compared to RCP4.5, the posterior distribution of ECS in the real world in Figure~\ref{fig:ecs-rcp85} is surprisingly different to Figure~\ref{fig:ecs} of the main text.
Under the RCP4.5 scenario, we estimated a median ECS for the real world of \SI{3.2}{\kelvin} with 90\,\% credible interval \SIrange{2.1}{5.1}{\kelvin}.
Under the RCP8.5 scenario we estimate a median ECS of \SI{3.2}{\kelvin} with 90\,\% credible interval \SIrange{2.1}{4.8}{\kelvin}.
As expected, there is little difference in the distribution of ECS because most of the information comes from the \abrupt experiment common to both estimates.

\subsection{Projections}

\begin{figure}[t]
  \includegraphics[scale=1]{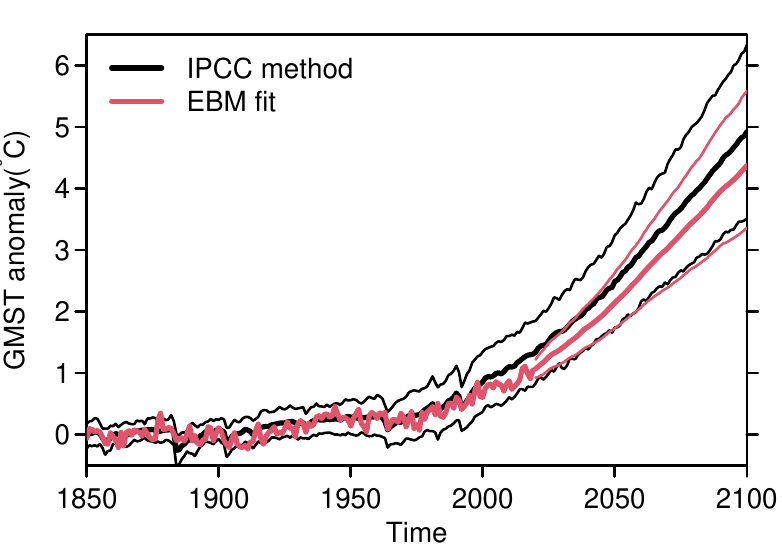}
  \caption{Global mean surface temperature anomaly projections. 
  Posterior predictive distribution for global mean surface temperature under  the RCP8.5 forcing scenario. 
  Thick lines indicate the posterior predictive means of the CMIP5 ensemble (black) and the real world (red). 
  Thin lines indicate marginal 90\,\% credible intervals. Grey lines are the individual CMIP5 model simulations.}
  \label{fig:projections-rcp85}
\end{figure}

Figure~\ref{fig:projections-rcp85} shows the projections under the RCP8.5 scenario for comparison with Figure~\ref{fig:projections} of the main text.
Once again, the projections are well constrained compared to the IPCC-style projections.

Table~\ref{tab:rcp85} compares our projections of the mean warming in the period 2081-2100 above the 1986-2005 average with other recent studies.
Similar to the RCP4.5 scenario, our median estimate of \SI{3.3}{\kelvin} is around \SI{0.2}{\kelvin} lower than other recent estimates that all project around \SI{3.5}{\kelvin} of warming by the end of the century.
Once again, our credible intervals are narrower than all except \citet{Strobach2020}.

We do not reproduce Figure~\ref{fig:probabilities} from the main text, since from Figure~\ref{fig:projections-rcp85} there is clearly little probability of remaining below \SI{1.5}{\kelvin} after around 2040, or below \SI{2.0}{\kelvin} after around 2060 under the stronger scenario.

\bibliographystyle{plainnat}
\bibliography{library}

\newpage

\begin{table}[h]
  \caption{Parameter estimates. The posterior means of the individual model parameters, the CMIP5 ensemble parameters and the observation parameters. Models highlighted in red were not included in the ensemble analysed in the main text}
  \begin{tabular}{lcccccccccccccc}
  \hline \hline
  Model & $\gamma$ & $C_1$ & $C_2$ & $C_3$ & $k_1$ & $k_2$ & $k_3$ & $\varepsilon$ & $\sigma_F$ & $\sigma_T$ & $F_C$ & $F_V$ & $\sigma_\delta$ \\
  \hline
{\color{red} ACCESS1-0}     & 2.28 & 4.27 & 12.4 & 78 & 0.74 & 2.71 & 1.08 & 1.49 & 0.66 & 0.46 & 3.31 & 18.2 & 0.033 \\
{\color{red} ACCESS1-3}     & 2.12 & 4.10 & 14.1 & 95 & 0.79 & 2.65 & 1.15 & 1.42 & 0.52 & 0.47 & 3.22 & 16.9 & 0.029 \\
BCC-CSM1.1    & 3.81 & 3.67 & 10.0 & 52 & 1.23 & 2.97 & 0.68 & 1.25 & 0.69 & 0.37 & 3.59 & 23.6 & 0.055 \\
{\color{red} BCC-CSM1.1(m)} & 3.13 & 3.86 & 9.7 & 47 & 1.29 & 2.38 & 0.63 & 1.30 & 0.88 & 0.41 & 3.82 & 23.0 & 0.054 \\
{\color{red} BNU-ESM }      & 2.40 & 3.82 & 10.6 & 86 & 0.97 & 1.91 & 0.64 & 1.06 & 0.76 & 0.62 & 3.88 & 20.6 & 0.039 \\
CanESM2       & 1.97 & 3.98 & 12.4 & 72 & 1.01 & 1.91 & 0.72 & 1.29 & 0.62 & 0.51 & 4.00 & 17.5 & 0.033 \\
CCSM4         & 2.64 & 4.27 & 12.8 & 76 & 1.30 & 2.36 & 1.18 & 1.35 & 0.65 & 0.52 & 4.06 & 23.0 & 0.041 \\
CNRM-CM5      & 3.37 & 3.61 & 10.3 & 90 & 1.14 & 2.49 & 0.61 & 0.98 & 0.56 & 0.43 & 3.68 & 21.3 & 0.046 \\
{\color{red} CSIRO-Mk3.6.0} & 1.91 & 4.13 & 13.3 & 69 & 0.62 & 2.62 & 1.11 & 1.63 & 0.73 & 0.51 & 3.08 & 16.9 & 0.030 \\
FGOALS-s2     & 2.04 & 4.62 & 11.5 & 129 & 0.90 & 1.84 & 1.02 & 1.15 & 0.75 & 0.65 & 3.93 & 20.2 & 0.033 \\
{\color{red} GFDL-CM3}      & 3.02 & 3.82 & 10.9 & 70 & 0.75 & 2.96 & 0.92 & 1.28 & 0.81 & 0.55 & 3.20 & 22.6 & 0.043 \\
GFDL-ESM2G    & 2.42 & 4.44 & 12.4 & 106 & 1.50 & 1.99 & 1.22 & 1.19 & 0.72 & 0.52 & 3.62 & 21.9 & 0.039 \\
{\color{red} GFDL-ESM2M}    & 2.52 & 4.45 & 13.2 & 105 & 1.49 & 2.12 & 1.52 & 1.21 & 0.79 & 0.65 & 3.69 & 24.5 & 0.040 \\
GISS-E2-R     & 2.72 & 5.24 & 15.2 & 134 & 1.83 & 2.50 & 2.15 & 1.39 & 0.45 & 0.38 & 4.20 & 22.8 & 0.037 \\
HadGEM2-ES    & 1.81 & 4.07 & 11.4 & 91 & 0.63 & 2.17 & 0.65 & 1.33 & 0.57 & 0.39 & 3.20 & 13.8 & 0.028 \\
{\color{red} IPSL-CM5A-LR}  & 2.43 & 3.75 & 12.0 & 88 & 0.76 & 2.54 & 0.77 & 1.20 & 0.59 & 0.48 & 3.29 & 18.1 & 0.034 \\
IPSL-CM5A-MR  & 2.37 & 3.90 & 12.1 & 93 & 0.76 & 2.55 & 0.76 & 1.23 & 0.51 & 0.43 & 3.43 & 17.0 & 0.032 \\
{\color{red} IPSL-CM5B-LR}  & 2.53 & 3.47 & 13.0 & 56 & 1.07 & 2.63 & 0.89 & 1.37 & 0.73 & 0.48 & 2.95 & 19.1 & 0.038 \\
{\color{red} MIROC-ESM}     & 2.20 & 5.01 & 11.8 & 134 & 0.80 & 2.25 & 1.09 & 1.26 & 0.56 & 0.48 & 4.22 & 19.2 & 0.032 \\
MIROC5        & 1.58 & 4.53 & 17.9 & 147 & 1.56 & 1.55 & 1.74 & 1.20 & 0.51 & 0.82 & 4.39 & 19.7 & 0.026 \\
MPI-ESM-LR    & 2.04 & 4.03 & 13.8 & 73 & 1.12 & 1.98 & 0.93 & 1.34 & 0.57 & 0.60 & 4.37 & 19.5 & 0.034 \\
{\color{red} MPI-ESM-MR}    & 1.95 & 4.18 & 13.7 & 75 & 1.15 & 1.96 & 0.90 & 1.38 & 0.54 & 0.50 & 4.32 & 18.1 & 0.032 \\
MRI-CGCM3     & 2.24 & 3.80 & 13.1 & 67 & 1.19 & 2.35 & 0.79 & 1.36 & 0.55 & 0.37 & 3.34 & 16.5 & 0.035 \\
NorESM1-M     & 1.91 & 4.77 & 14.7 & 104 & 1.09 & 2.27 & 1.56 & 1.51 & 0.59 & 0.46 & 3.50 & 17.6 & 0.029 \\
\hline
Ensemble      & 2.40 & 4.17 & 12.6 & 90 & 1.08 & 2.33 & 1.04 & 1.30 & 0.64 & 0.50 & 3.68 & 19.7 & 0.036 \\
\hline
Observations  & 2.11 & 4.34 & 14.1 & 95 & 1.02 & 2.42 & 1.17 & 1.45 & 0.56 & 0.43 & 3.49 & 16.8 & 0.031 \\
\hline
  \end{tabular}
  \label{tab:parameters-all}
\end{table}

\begin{table}[h]
  \caption{Parameter estimates when fitted to the RCP8.5 future scenario. The posterior means of the individual model parameters, the CMIP5 ensemble parameters and the observation parameters.}
  \begin{tabular}{lcccccccccccccc}
  \hline \hline
  Model & $\gamma$ & $C_1$ & $C_2$ & $C_3$ & $k_1$ & $k_2$ & $k_3$ & $\varepsilon$ & $\sigma_F$ & $\sigma_T$ & $F_C$ & $F_V$ & $\sigma_\delta$ \\
  \hline
BCC-CSM1.1   & 3.53 & 4.00 & 9.5 & 53 & 1.22 & 2.64 & 0.71 & 1.29 & 0.63 & 0.37 & 3.59 & 23.7 & 0.048 \\
CanESM2      & 2.13 & 4.01 & 11.4 & 72 & 0.98 & 2.05 & 0.73 & 1.27 & 0.65 & 0.57 & 3.91 & 18.8 & 0.038 \\
CCSM4        & 2.61 & 4.37 & 13.9 & 77 & 1.29 & 2.02 & 1.22 & 1.37 & 0.64 & 0.50 & 4.05 & 23.0 & 0.040 \\
CNRM-CM5     & 2.96 & 3.67 & 10.9 & 81 & 1.14 & 2.41 & 0.65 & 1.00 & 0.51 & 0.44 & 3.70 & 20.5 & 0.047 \\
FGOALS-s2    & 1.94 & 4.55 & 14.4 & 122 & 0.84 & 1.59 & 1.30 & 1.34 & 0.72 & 0.64 & 3.99 & 19.5 & 0.034 \\
GFDL-ESM2G   & 2.31 & 4.63 & 16.6 & 95 & 1.50 & 1.76 & 1.47 & 1.34 & 0.70 & 0.54 & 3.71 & 22.6 & 0.038 \\
GISS-E2-R    & 3.04 & 5.06 & 24.6 & 120 & 1.80 & 2.03 & 2.99 & 1.39 & 0.47 & 0.36 & 4.12 & 23.4 & 0.037 \\
HadGEM2-ES   & 2.16 & 4.26 & 10.1 & 89 & 0.63 & 2.33 & 0.66 & 1.31 & 0.61 & 0.42 & 3.18 & 14.9 & 0.033 \\
IPSL-CM5A-MR & 2.34 & 4.15 & 12.0 & 95 & 0.78 & 2.36 & 0.78 & 1.21 & 0.53 & 0.42 & 3.46 & 16.3 & 0.035 \\
MIROC5       & 1.58 & 4.39 & 23.6 & 134 & 1.59 & 1.53 & 1.68 & 1.15 & 0.52 & 0.75 & 4.40 & 19.0 & 0.034 \\
MPI-ESM-LR   & 2.16 & 4.05 & 13.1 & 76 & 1.12 & 1.93 & 0.91 & 1.29 & 0.61 & 0.61 & 4.31 & 20.6 & 0.039 \\
MRI-CGCM3    & 2.38 & 4.16 & 12.6 & 67 & 1.22 & 2.58 & 0.74 & 1.25 & 0.56 & 0.41 & 3.34 & 17.0 & 0.037 \\
NorESM1-M    & 1.97 & 4.94 & 17.9 & 106 & 1.12 & 2.05 & 1.50 & 1.43 & 0.61 & 0.46 & 3.48 & 16.3 & 0.030 \\
\hline
Ensemble     & 2.41 & 4.34 & 14.8 & 92 & 1.18 & 2.11 & 1.20 & 1.28 & 0.60 & 0.50 & 3.79 & 19.7 & 0.038 \\
\hline
Observations & 2.22 & 4.46 & 15.1 & 95 & 1.12 & 2.31 & 1.14 & 1.31 & 0.57 & 0.45 & 3.49 & 16.9 & 0.034 \\
\hline
  \end{tabular}
  \label{tab:parameters-rcp85}
\end{table}

\begin{table}[ht]
  \caption{Estimates of future warming under RCP8.5 scenario averaged over 2081--2100. Note that the IPCC and \citet{Sherwood2020} intervals are 66\,\% intervals, not 90\,\% intervals.}
  \begin{tabular}{lllll}
  \hline \hline
  Study & Ensemble & Reference & Median & \SIrange{5}{95}{\percent} \\
  \hline
  IPCC & CMIP5 & 1986--2005 & \SI{3.7}{\kelvin} & \SIrange{2.6}{4.8}{\kelvin} \\
  \citet{Sherwood2020} & CMIP5 & 1986--2005 & \SI{3.5}{\kelvin} & \SIrange{3.0}{4.2}{\kelvin} \\
  \citet{Strobach2020} & CMIP5 & 1986--2005 & \SI{3.5}{\kelvin} & \SIrange{3.4}{3.9}{\kelvin} \\
  \citet{Tokarska2020} & CMIP6 & 1995--2014 & \SI{3.4}{\kelvin} & \SIrange{2.3}{4.6}{\kelvin} \\
  This study & CMIP5 & 1986--2005 & \SI{3.3}{\kelvin} & \SIrange{2.5}{4.4}{\kelvin} \\
  \hline
  \end{tabular}
  \label{tab:rcp85}
\end{table}

\newpage

